\documentclass[letterpaper]{article}
\usepackage[ margin= 0.993 in]{geometry}
\usepackage{graphicx}
\usepackage[english]{babel}
\usepackage{amsmath,amsthm,amsfonts,amssymb,graphicx,fullpage,listings, latexsym}
\usepackage{bbm}
\usepackage[labelfont=bf]{caption}
\usepackage{xcolor}
\usepackage{subfigure}
\usepackage{float}
\usepackage{url}
\usepackage[colorlinks=true,pdfborder={0 0 0}]{hyperref}
\usepackage{algorithm}
\usepackage{algpseudocode}
\usepackage{comment}
\usepackage{booktabs}
\usepackage{titlesec} 

\usepackage{hyperref}
\newcommand{\deln}{\delta_0}
\newcommand{\mut}{\hat{\mu}_t}
\newcommand{\sigt}{\hat{\sigma}_t}

\theoremstyle{plain}
\newtheorem{theorem}{Theorem}[section]
\newtheorem{lemma}{Lemma}[section]
\newtheorem{corollary}{Corollary}[section]
\newtheorem{remark}{Remark}[section]

\long\def\symbolfootnote[#1]#2{\begingroup
\def\thefootnote{\fnsymbol{footnote}}\footnote[#1]{#2}\endgroup}

\begin{document}
\title{ Data-Adaptive Symmetric CUSUM for \\ Sequential Change Detection }
\author{
Nauman Ahad, Mark A. Davenport, and Yao Xie}
\maketitle



\begin{abstract}
   Detecting change points sequentially in a streaming setting, especially when both the mean and the variance of the signal can change, is often a challenging task. A key difficulty in this context often involves setting an appropriate detection threshold, which for many standard change statistics may need to be tuned depending on the pre-change and post-change distributions.  This presents a challenge in a sequential change detection setting when a signal switches between multiple distributions. For example, consider a signal where change points are indicated by increases/decreases in the mean and variance of the signal. In this context, we would like to be able to compare our change statistic to a fixed threshold that will be symmetric to either increases or decreases in the mean and variance. Unfortunately, change point detection schemes that use the log-likelihood ratio, such as CUSUM and GLR, are quick to react to changes but are not symmetric when both the mean and the variance of the signal change. This makes it difficult to set a single threshold to detect multiple change points sequentially in a streaming setting. We propose a modified version of CUSUM that we call Data-Adaptive Symmetric CUSUM (DAS-CUSUM). The DAS-CUSUM change point detection procedure is symmetric for changes between distributions, making it suitable to set a single threshold to detect multiple change points sequentially in a streaming setting. We provide results that relate to the expected detection delay and average run length for our proposed procedure. Extensive simulations are used to validate these results. Experiments on real-world data further show the utility of using DAS-CUSUM over both CUSUM and GLR. 
\end{abstract}
{\let\thefootnote\relax\footnote{{N.Ahad and M.A.Davenport are with the School of Electrical and Computer Engineering, Georgia Tech, Atlanta,  GA, 30302, USA.
Y.Xie is with the School of Industrial and Systems Engineering, Georgia Tech, Atlanta, GA, 30302, USA. The work of N. Ahad and M. Davenport was supported, in part, by NSF grants CCF-2107455 and DMS-2134037, NIH grant R01AG056255, and gifts from the Alfred P. Sloan Foundation and Coulter Foundation. The work of Y. Xie was supported, in part, by an NSF CAREER grant CCF-1650913, and NSF grants DMS-2134037, CMMI-2015787, DMS-1938106, and DMS-1830210. \\ E-mails:\texttt{ nahad3@gatech.edu, mdav@gatech.edu, yao.xie@isye.gatech.edu} }}}

\section{Introduction}\label{sec:intro}

For a sequence of observations $x_1, \ldots, x_t$, the goal of change point detection is to detect whether there exists an instance $n_c$ such that $x_1, \ldots, x_{n_c -1}$ are generated according to a different distribution than $x_{n_c}, \ldots, x_t$, and if so, estimating $n_c$. This is typically accomplished by computing a simple change statistic based on the log-likelihood ratio, which can be compared to a threshold to detect changes or optimized to estimate $n_c$.  Sequential change point detection involves sequentially detecting multiple changes in streaming data. Many real-world world applications require sequential detection of change points within streaming signals. Healthcare, communication, and finance are just a few areas where sequential change detection is widely used \cite{yang2006adaptive, lai2008quickest, al2006surface}.  An extended discussion of applications of change point detection can be found in \cite{aminikhanghahi2017survey}. 

Despite being devised more than half a century ago, the CUSUM statistic is still one of the most popular methods for detecting change points \cite{page1954continuous}. 
This is chiefly due to two reasons. First, it has a simple recursive implementation which makes it computationally efficient to apply. Second, it has been shown to be optimal in minimizing the detection delay for a given false alarm rate~\cite{lorden1971procedures}. However, computing the CUSUM statistic requires complete knowledge of both the pre-change and post-change distributions. This is not feasible in many real-world scenarios where the post-change distribution can be unknown. In such settings, a more common approach is to use the GLR statistic, which involves estimating the post-change distribution for all possible change points~\cite{siegmund1995using}. Both the CUSUM and GLR statistics leverage the log-likelihood ratio for the known/estimated pre- and post-change distributions. 

Most work on change point detection has focused on identifying a \emph{single} change point in the quickest possible manner. Though this has been useful for some applications, especially those that monitor a process for abnormal behavior 
 such as machine fault detection and network intrusion detection, many modern applications require the detection of \emph{multiple} change points sequentially in streaming data. In sequential change point detection, the detection procedure must be restarted and continued after each change point is detected, resulting in multiple change points being detected. Examples of such settings include segmentation of signals for activity recognition where change points are used to identify transitions from one activity to another in a streaming setting~\cite{aminikhanghahi2017using}. In such settings, the pre-change and post-change distributions themselves change after each change point and cannot be assumed to be known \emph{a priori}.  This presents a significant challenge to most standard change detection approaches because the detection threshold must be set without any knowledge of these distributions (with the threshold typically being fixed in advance and held constant throughout the procedure).

The machine learning community has been addressing this problem of identifying multiple change points in data streams \cite{liu2013change}. Such works show that procedures employed to detect change points should be \emph{symmetric.} This means the magnitude of a change from a distribution $\theta_0$ to a distribution $\theta_1$ should be the same for a change from $\theta_1$ to $\theta_0$. Using a procedure that has a similar power in detecting such changes makes it easy to select a threshold for detecting multiple changes sequentially. Statistics such as the GLR and CUSUM are not symmetric when distribution changes involve a change in variance. This makes it difficult to use these in detecting multiple changes.  

In this work, we present an adaptive symmetric version of CUSUM  that we call Data-Adaptive Symmetric CUSUM (DAS-CUSUM). DAS-CUSUM uses a window to estimate the post-change distribution and employs a symmetric change statistic to make it easier to select a fixed threshold to detect multiple change points in streaming data. We provide theoretical results for our proposed method that relate the expected detection delay (EDD) (average delay in detecting true changes) to the average run length (ARL) (average time until a false alarm occurs). 
 
The rest of the paper is organized as follows. After reviewing related literature in Section \ref{sec: lit review}, we formalize the change detection problem in Section \ref{sec: prob set up}  and further motivate the need to have a symmetric change statistic for detecting multiple changes. Section \ref{sec: proposed method} provides a  description of the proposed procedure. Theoretical results that relate EDD versus ARL are described in Section \ref{sec: theoratical}, where a sketch of the related proofs is also given. Section \ref{sec: simulations} contains simulations that empirically validate the theoretical results in a practical setting. Experiments on real-world data are summarized in Section \ref{sec: real data}.
    
\section{Related work}\label{sec: lit review}

The CUSUM statistic is known for being asymptotically optimal in minimizing the maximum average detection delay as the average time to false alarm reaches infinity  \cite{lorden1971procedures}. CUSUM was later shown to be optimal in minimizing the expected detection delay for a provided (non-asymptotic) expected time to false alarm \cite{moustakides1986optimal}.   There has been extensive work done to further investigate and generalize the optimality property of CUSUM. These results, however hold when both pre-change and post-change distribution are completely known. A summary of such work can be found in \cite{veeravalli2014quickest}. A two-sided CUSUM test can be used to detect either an increase or decrease in mean \cite{granjon:hal-00914697}, but this approach still assumes a fixed and known variance.  When the post-change distribution is unknown, the generalized log-likelihood ratio test (GLR) can be used by estimating both the change location and the post-change distribution through maximum likelihood estimation.  However, CUSUM, GLR, and their variants are often used to detect only a \emph{single} change point \cite{tartakovsky2006novel}. The few works that do use these methods to detect multiple changes do so by only detecting changes in the mean of normally distributed data \cite{bodenham2017continuous,fathy2018online}. It is more challenging to detect multiple changes when both the mean and the variance of a signal change. There is limited prior work that detects joint changes in both the mean and the variance of the signal  \cite{hawkins2005statistical}, however, this has not been considered in the context of detecting multiple changes.

Recently, there has been increasing interest in the machine learning community to detect multiple change points sequentially within streaming data \cite{kifer2004detecting, liu2013change, alippi2016change,chang2018kernel}. Most of these methods use non-parametric change statistics, which are symmetrical.  This means that the magnitude of the change statistic for a change from $\theta_0$ to $\theta_1$ is equivalent in magnitude for a change from $\theta_1$ to $\theta_0$. The need for this symmetrical statistic was noted by \cite{liu2013change}, who use a symmetric KL-divergence to detect multiple changes within streaming data where both the mean and variance of the normally distributed signal are changing. The symmetric statistic makes it easy to set a single detection threshold before the procedure is started to detect multiple changes within streaming data. At each time instance, a  pre-change distribution is estimated using a ``past window,'' and the post-change distribution is estimated using a ``future window.''  These methods, however, do not incorporate data samples directly. These samples are incorporated through estimates of the distribution, which makes these methods slow to react to changes.  None of these methods characterize the relationship between detection delay and false alarm rate.

The need to  use symmetric statistics for change detection was also earlier noticed in \cite{basseville1983sequential, andre1988new,gustafsson2000adaptive},  where the authors noted the asymmetry in change statistics when there are changes in both mean and variance. These works used a log-likelihood ratio with a drift term to make the expected value of the change statistic symmetric under the post-change distribution. However, this drift term meant that the expected value of the statistic is zero under the pre-change distribution, which can lead to more false positives. A slightly modified version of this technique was mentioned in \cite{basseville1993detection},  where false alarm rates were reduced by adding a fixed drift term which made the expected value of the statistic negative under the pre-change distribution. However, no details were provided about setting this drift term. These methods also provided no characterization of the relationship between detection delay and false alarm rate.

In this work, we investigate a suitable choice for this fixed drift to make the statistic symmetric under the post-change distribution while also ensuring the expectation is negative under the pre-change distribution. 
Our proposed change detection procedure provides a symmetric change statistic for different families of probability distributions, however, the theoretical results relating detection delay and false alarm rate consider the more restricted setting of i.i.d.\ univariate normally distributed data.

 \section{Problem statement}\label{sec: prob set up}

Change points are instances in a signal where the underlying distribution of data changes, e.g., the parameters of the signal generating distribution change from $\theta_0$ to $\theta_1$. Most change point detection methods rely on hypothesis tests based on the log-likelihood ratio. Specifically, suppose we are given observations $x_0, \ldots, x_t$ of a time series $X$. We will assume that each element $x_i$ is drawn independently from a distribution $f_\theta$ where $\theta$ represents some (possibly changing) parameters. To detect a change we compare the null hypothesis ($\mathcal{H}_0$) that all $x_i$ are drawn according to $f_{\theta_0}$ for some (known) $\theta_0$ to the alternate hypothesis ($\mathcal{H}_1$) that the time series distribution changes from $f_{\theta_0}$ to $f_{\theta_1}$, at time $n_c$, for some $\theta_1 \neq \theta_0$. 

The likelihood of $X$ under these two hypotheses is given by:
\begin{align*} 
\mathcal{L}(\mathcal{H}_0 | X) & = \prod_{i=1}^{t} f_{\theta_0}(x_i) \\
\mathcal{L}(\mathcal{H}_1 | X) &= \prod_{i=1}^{{n_c}-1} f_{\theta_0}(x_i)\prod_{i=n_c}^{t} f_{\theta_1}(x_i).
\end{align*}
By computing the likelihood ratio  and taking the logarithm, we obtain the likelihood-ratio statistic at instance  $t$ for a change at $n_c$:
\begin{equation*}
 \ell^t_{n_c} = \sum_{{i=n_c}}^t \log \frac{f_{\theta_1}(x_i)}{f_{\theta_0}(x_i)}.
\end{equation*}

Since the location of the change point $n_c$ is unknown, the maximum over all possible change point locations is taken to compute the change statistic at instance $t$:

\begin{equation}
    \ell^t = \mathop{\max}_{1 < n_c < t} \ell_{n_c}^t.
    \label{eq: LR_max}
\end{equation}

A change point is detected the first time the change statistic $\ell^t$ is greater than a specified threshold $b$. For a sequence of i.i.d.\ random variables, the sum of the log-likelihood probability ratio between distributions $\theta_1$ and $\theta_0$ satisfies an intuitive property: if a sequence is generated through a post-change distribution, the expected value of this sum should be positive. If the sequence is generated through the pre-change distribution, this sum should be negative. Concretely speaking, if we represent his log-likelihood ratio as
\begin{equation*}
         \ell^t_{1}  = \sum_{{i=1}}^t \log \frac{f_{\theta_1}(x_i)}{f_{\theta_0}(x_i)} ,
\end{equation*}
then  $\mathbb{E}_{\theta_1}(\ell^t_1) > 0 $ and $\mathbb{E}_{\theta_0}(\ell^t_1) < 0 $. In~\eqref{eq: LR_max}, we are maximizing over $n_c$ to find the maximum log-likelihood ratio. Instead of maximizing~\eqref{eq: LR_max} with respect to $n_c$, we can also maximize the log-likelihood ratio by minimizing, over $n_c$, the expression:

\begin{equation}
    \ell^t = \sum_{{i=1}}^t \log \frac{f_{\theta_1}(x_i)}{f_{\theta_0}(x_i)} - \mathop{\min}_{1<n_c<t} \sum_{{i=1}}^{n_c} \log \frac{f_{\theta_1}(x_i)}{f_{\theta_0}(x_i)}.
        \label{eq: LR_min}
\end{equation}

The CUSUM statistic \cite{page1954continuous} provides a computationally attractive recursive implementation of the test in~\eqref{eq: LR_min}. It assumes both pre-change parameters $\theta_0$ and post-change $\theta_1$ distribution parameters are known. 
In such a setting,  a recursive implementation of~\eqref{eq: LR_min} can be obtained as shown below in~\eqref{eq: CUSUM} :
\begin{align}
     S_t &=  \left( S_{t-1}  + \log\frac{f_{\theta_1}(x_t)}{f_{\theta_0}(x_t)} \right)^+
     \label{eq: CUSUM},
\end{align}
where $( S_{t-1} )^+ = \max(0, S_{t-1})$ and $S_0 = 0$.  A change is detected at the first instance, $n_c$, where the corresponding change statistic $S_t$ is greater than a set threshold $b$. This is made concrete in the equation below:
\begin{equation*}
    n_c = \inf( t > 0 : S_t > b ) .
\end{equation*}

The post-change distribution is often unknown in real-world settings. In such cases, the GLR \cite{siegmund1995using} can be used to obtain the change statistic $\ell_t$. GLR maximizes the change statistic in~\eqref{eq: LR_GLR} over both the post-change distribution, $\theta_t$ at instance $t$, as well as the change instance $n_c$. 
Let \[
\ell^t_{n_c}:=\max_{\theta_t} \sum_{{i={n_c}}}^t \log \frac{f_{\theta_t}(x_i)}{f_{\theta_0}(x_i)}.
\]
Define
\begin{equation}
    \ell^t = \max_{1 < n_c < t}\ell_t^{n_c}.
        \label{eq: LR_GLR}
\end{equation}
This is done by first choosing a possible change instance, $n_c$, and finding the maximum likelihood estimate (MLE) $\hat{\theta}_t^{n_c}$ for~\eqref{eq: LR_GLR}. This MLE estimate is used to obtain a possible change statistic, $\ell_{n_c}^t$, corresponding to a change at $n_c$. This is repeated for all possible change instances before $t$, and the maximum of these is taken as the change statistic $\ell^t$ at time $t$.

Once the change statistic, $\ell_t$, crosses the threshold $b$, a change is detected, and the corresponding post-change estimate $\hat{\theta}_t^{n_c}$ is used as the new pre-change estimate $\theta_0$ and the sequential change point detection procedure is repeated to detect the next change. This way \emph{multiple} change points are detected.
It is important to point out that the GLR procedure is non-recursive and can be computationally expensive to run.

\subsection{Asymmetry of log-likelihood ratio}
\label{sec: Asym}
\begin{figure}
    \centering
 \subfigure[$\mathcal{N}(1, 1) \rightarrow \mathcal{N}(10, 3)$  ]{\label{fig: asym_int a}\includegraphics[width=60mm]{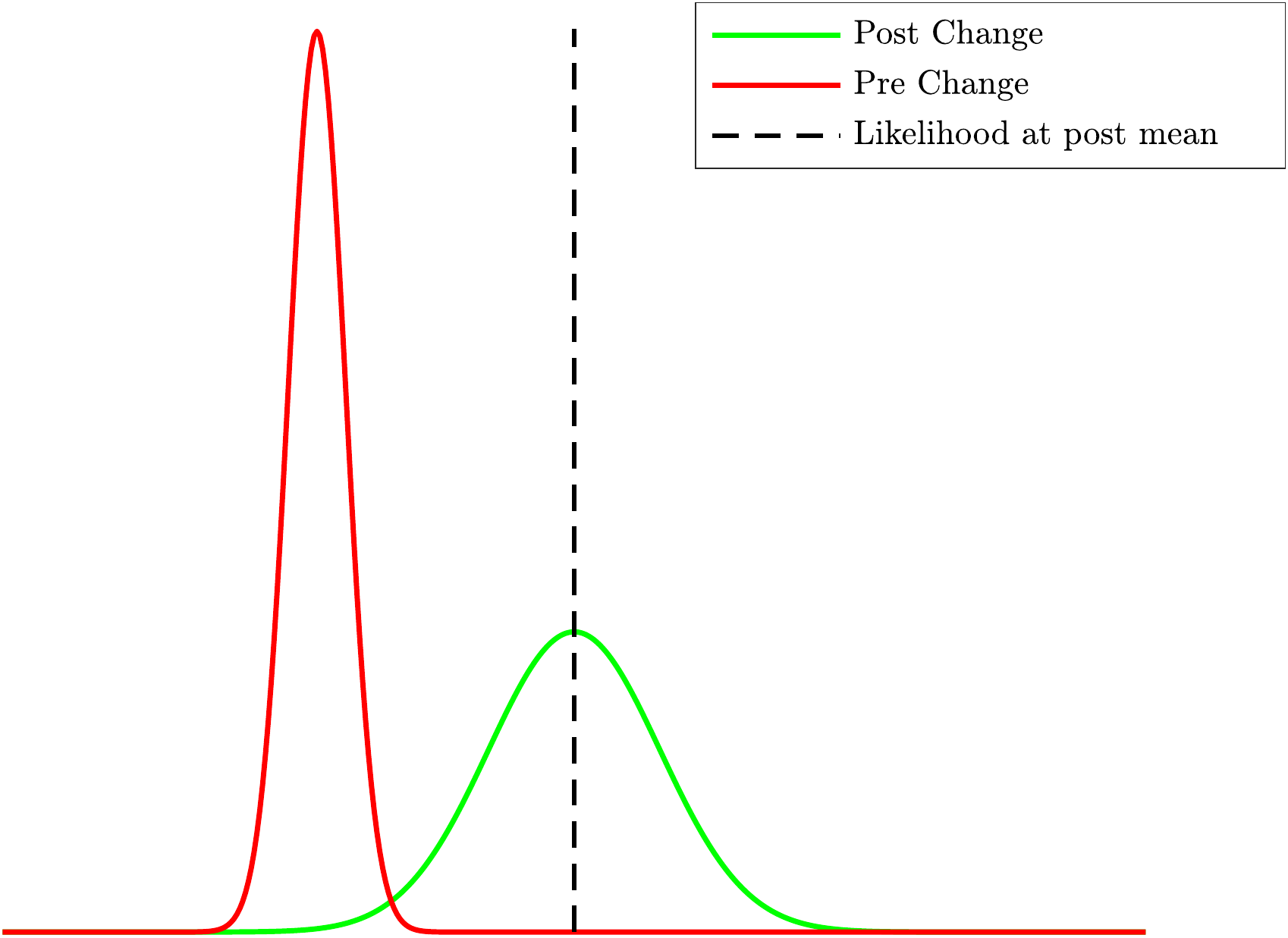}}
\subfigure[$\mathcal{N}(10, 3) \rightarrow \mathcal{N}(1, 1)$ ]{\label{fig: asym_int b}\includegraphics[width=60mm]{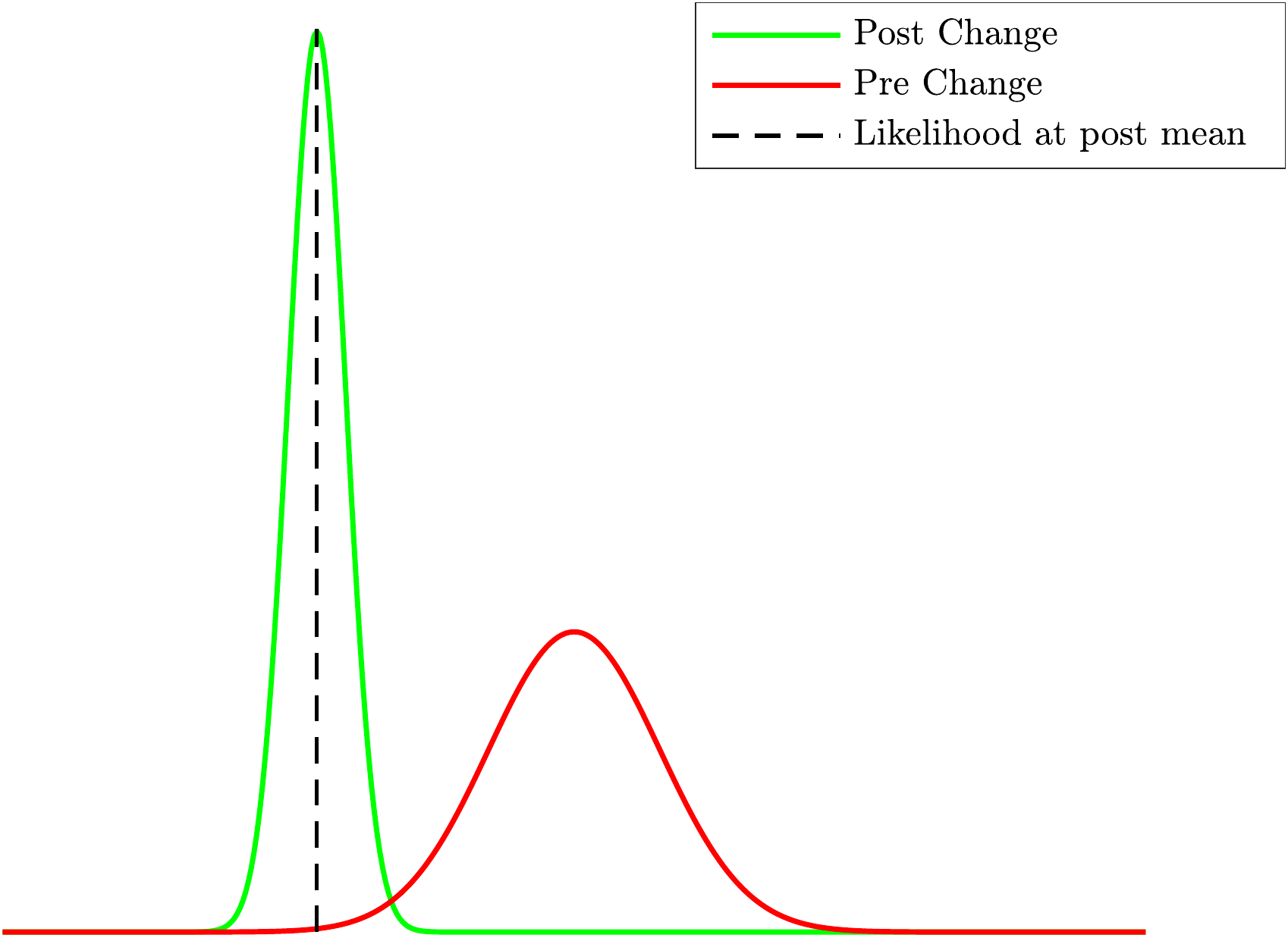}}  
   \caption{Joint changes in mean variance lead to asymmetric Likelihood Ratios. In Figure \ref{fig: asym_int a}, the pre-change likelihood is in the tail, leading to a large likelihood ratio. In Figure \ref{fig: asym_int b}, the post-change likelihood is higher than it is in Figure \ref{fig: asym_int a}, leading to a relatively smaller likelihood ratio}
   \label{Fig: Asymmetry}
\end{figure}

The log-likelihood ratio statistic, employed by both GLR and CUSUM, is quick to react to changes but is an asymmetric statistic for detecting joint changes in mean and variance.
Figure \ref{Fig: Asymmetry} illustrates this asymmetry. This difference gets more pronounced when one of the two distributions has a much smaller variance.  

\begin{figure}
    \centering
 \subfigure[Missed change point]{\label{fig: GLR asym real missed}\includegraphics[height=.55\textwidth]{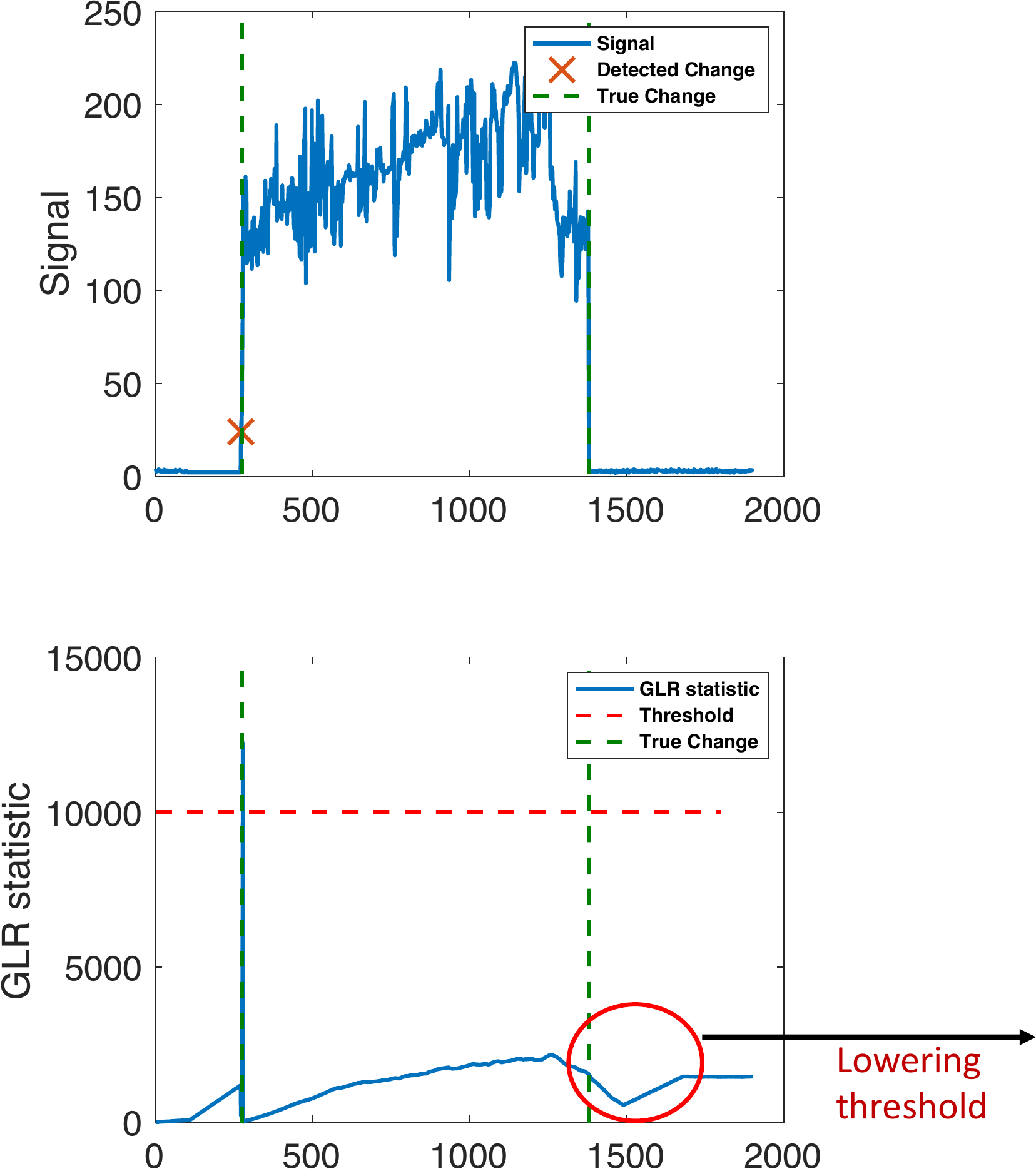}}
\subfigure[False change point]{\label{fig: asym real GLR false}\includegraphics[height=.55\textwidth]{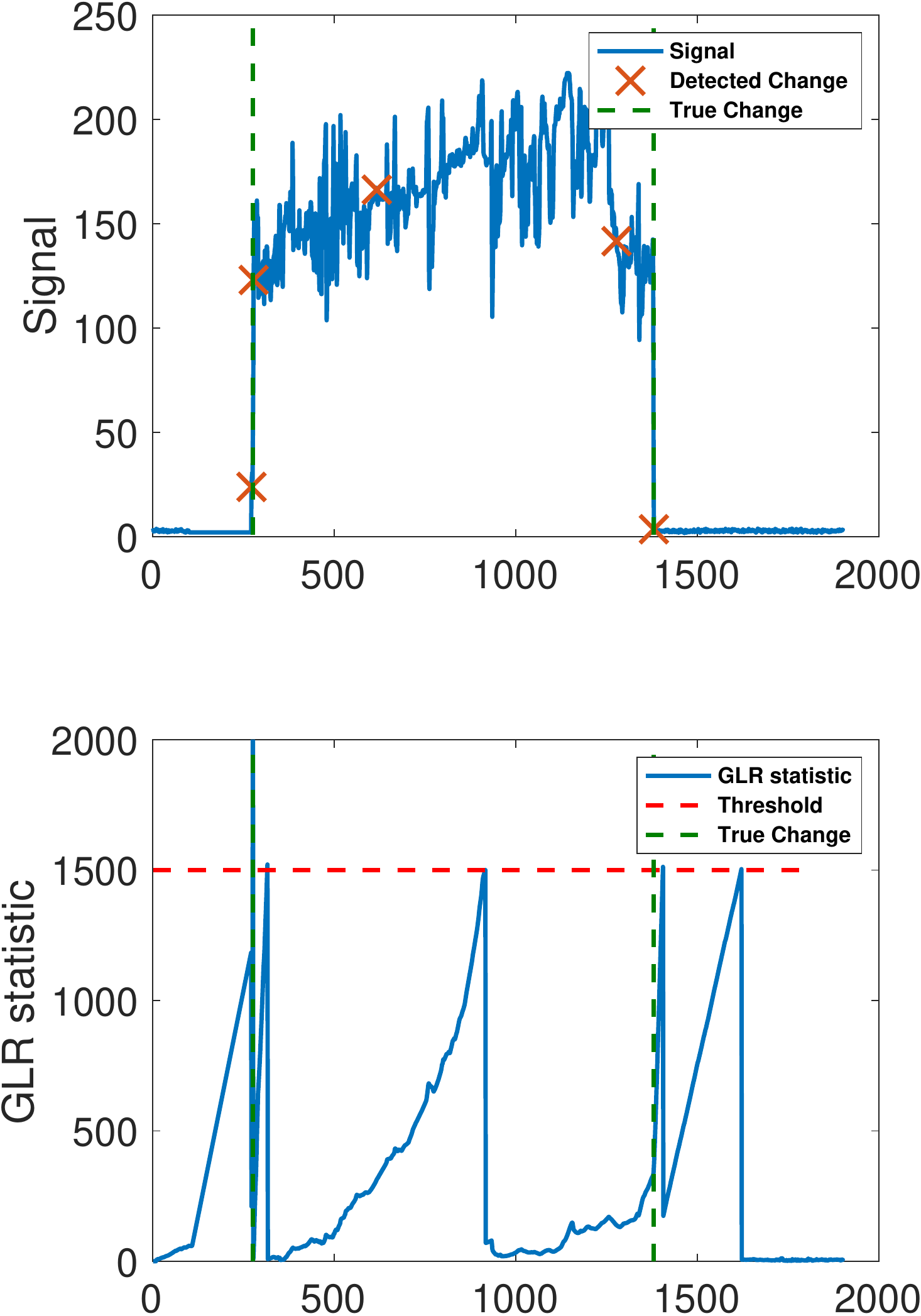}}  
   \caption{Joint changes in mean variance lead to asymmetric likelihood ratio. In Figure \ref{fig: GLR asym real missed}, the likelihood ratio (in GLR) for the second change is much smaller than the likelihood for the first change. This can lead to a missed change point when the detection threshold is set to be large. When the detection threshold is lowered to detect this missed change point, many false change points are detected, which can be seen in Figure \ref{fig: asym real GLR false}.}
   \label{Fig: Asymmetry real data}
\end{figure}

 Figure \ref{Fig: Asymmetry real data} shows a real-world example where this asymmetry makes it difficult for GLR to detect multiple change points.  The log-likelihood ratio for the first change point is much larger than the log-likelihood ratio for the second change point. This makes it difficult to set a detection threshold \emph{a priori} to detect multiple change points in a streaming data setting. In the first figure, the fixed detection threshold results in missing the second change point, which has a much smaller statistic. A reduction in the detection threshold leads to many false change point detections, which can be seen in the second figure.
 
\section{Data-adaptive symmetric CUSUM (DAS-CUSUM)}
\label{sec: proposed method}

\subsection{Adaptive post-change estimation}

\begin{figure}%
\centering
\includegraphics[width=0.8\textwidth]{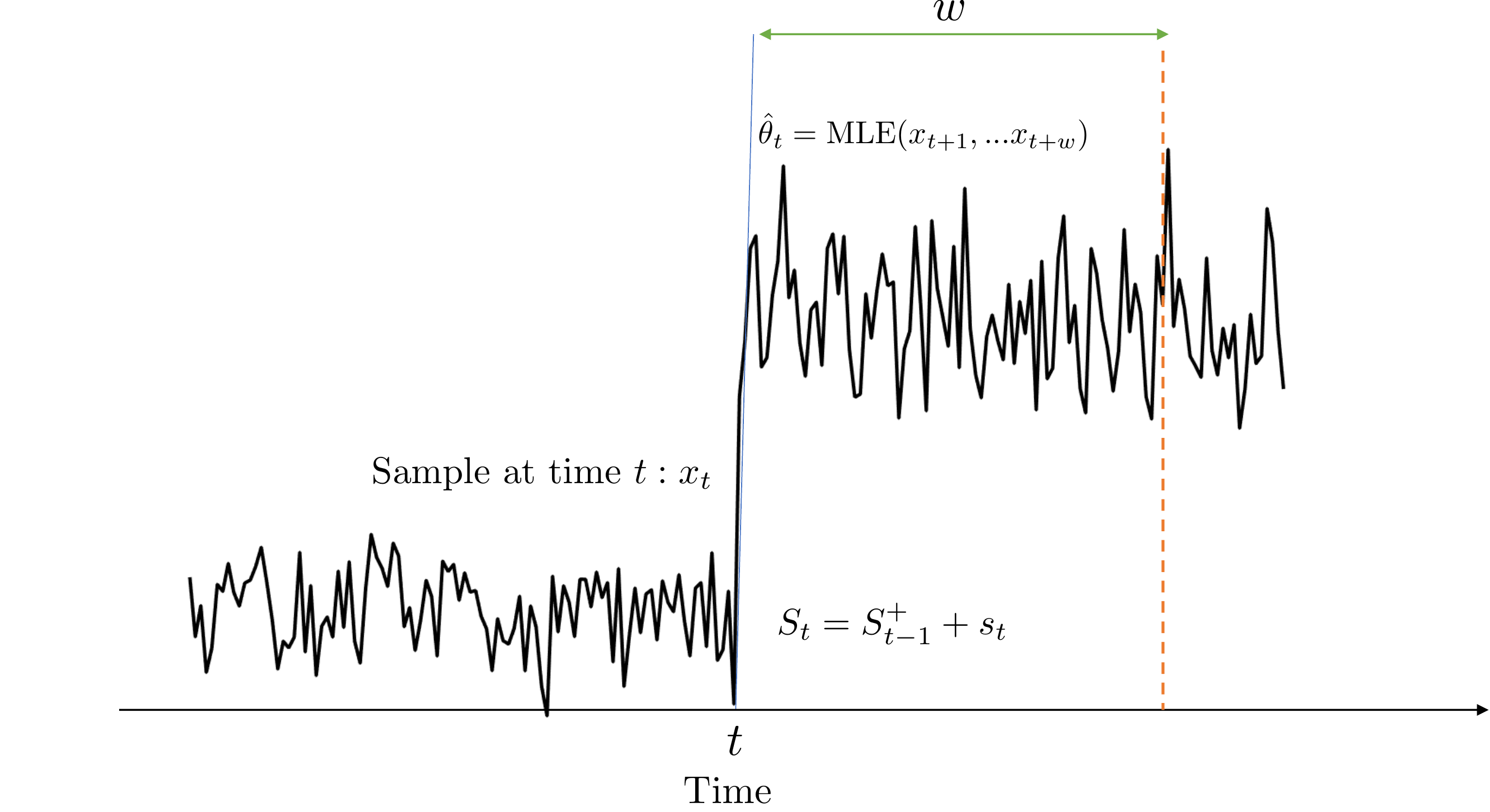}%
\caption{Adaptive version of CUSUM. Using a ``future'' window to estimate post-change parameters $\hat{\theta}_t$, which could be used in place of post-change distribution $\theta_1$ for CUSUM update.}
\label{fig: Apt CUSUM update}
\end{figure}

When the post-change distribution is not known, another way to estimate the post-change distribution is to use a window of size $w$ to estimate the post-change parameters $\hat{\theta}_t$ at time $t$ for the CUSUM statistic $S_t$. The same approach is used in \cite{xie2018first} where a window is used to estimate post-change distribution change distribution for subspace change detection. Figure \ref{fig: Apt CUSUM update} shows how this update is done.

For normally distributed i.i.d.\ data, the post-change distribution estimate $\hat{\theta}_t = ( \hat{\mu}_t, \hat{\sigma}_t^2 )$ at time $t$ is calculated via:
%
\[
    \hat{\mu}_t = \sum_{i=t+1}^{t+w} x_i, \quad
     \hat{\sigma}_t^2 = \sum_{i=t+1}^{t+w} \frac{1}{w} ( x_i -\hat{\mu}_t  )^2.
\]

Using ``future'' samples to calculate post-change estimates  $\hat{\theta}_t$ may seem unreasonable at first, but detection decisions can be delayed by $w$ samples so that data is available for calculating these estimates (provided, of course, that $w$ is not excessively large). These estimates can be substituted for $\theta_1$ in~\eqref{eq: CUSUM} to obtain an adaptive form of CUSUM where the post-change distribution is estimated.
Such estimates are also independent of the change statistic $S_t$.  In comparison to GLR, adaptive CUSUM leads to a more computationally efficient method for detecting change points when the post-change distribution is unknown. CUSUM has been extensively studied to develop tools that characterize the detection average run length (ARL), which is the average time till false detection under the pre-change distribution, and the expected detection delay (EDD), which is the expected time till true detection under the post-change distribution. Adaptive CUSUM can utilize the same tools to characterize the ARL and EDD performance.

\subsection{Proposed procedure} 
\label{sec: proposed statistic}
As discussed in Section \ref{sec: Asym}, the log-likelihood ratio test is asymmetric for changes between two distributions having different variances. This makes it difficult to select a single threshold for adaptive CUSUM to detect multiple changes.

To solve this problem, we introduce a symmetric version of adaptive CUSUM that we call DAS-CUSUM. 
The DAS-CUSUM based change detection procedure is given below:
\begin{align}
    S_t &= ( S_{t-1} )^+ + s_t \nonumber \\
    & = ( S_{t-1} )^+ + \log{\frac{f_{\hat{\theta}_t}(x_i)}{f_{\theta_0}(x_i)}} - \mathbb{E}_{\theta_0}\big(\log\frac{f_{\hat{\theta}_t }(x_i)}{f_{\theta_0}(x_i)}\big) - v \nonumber\\
    &= ( S_{t-1} )^+  + \log{\frac{f_{\hat{\theta}_t}(x_i)}{f_{\theta_0}(x_i)}} + D_{\rm KL}(f_{\theta_0}(x)  , f_{\hat{\theta}_t}(x) )   - v.
    \label{eq: DAS-CUSUM statistic indicator}
 \end{align}
Here, the incremental update statistic is given by:
 \begin{equation}
    \label{eq: DAS-CUSUM update}
    s_t =  \log{\frac{f_{\hat{\theta}_t}(x_i)}{f_{\theta_0}(x_i)}} + D_{KL}(f_{\theta_0}(x)  , f_{\hat{\theta}_t}(x) )  - v,
\end{equation}
where $D_{\rm KL}(f_{\theta_0}(x), f_{\hat{\theta}_t}(x) ))$ represents the term to ensure symmetry of the statistic (for either ``direction'' of change), and $v >0$ is the drift term. 

Compared to the incremental update for CUSUM, which only contains the log-likelihood ratio, the DAS-CUSUM update statistic contains two additional terms, which can be seen in~\eqref{eq: DAS-CUSUM update}. The first of these terms is a KL divergence which makes the incremental statistic almost symmetric under the post-change distribution. When $v$ is much smaller than the sum of two divergences, the expected value is almost the same for changes from distribution $\theta_0$ to $\theta_1$ and vice versa:
\begin{equation}
    \mathbb{E}_{\theta_1}[s_t] = D_{\rm KL}(\theta_1,\theta_0) + D_{\rm KL}(\theta_0 , \theta_1) - v.
    \label{eq: expect under postchange}
\end{equation}
The second of these additional terms, $v$, is a drift term that makes the expectation of the incremental statistic negative under the pre-change distribution. This allows our proposed statistic to match the property of CUSUM, which requires that the incremental statistic should be negative under the pre-change distribution to avoid detecting false change points, i.e., 
\[
    \mathbb{E}_{\theta_0}[s_t] =  -v.
\]

\subsection{Practical implementation}

Algorithm 1  shows how to implement DAS-CUSUM for detecting multiple change points. This algorithm uses values for the window size $w^*$ and drift term $v^*$, which are based on theoretical results presented in Section \ref{sec: theoratical}. These results, however, require complete knowledge of the post-change distribution $\theta_1$ to compute the symmetric KL divergence $s$ which is needed to compute the desired values for  $w^*$ and $v^*$. Since this post-change distribution is unknown, we can set a minimum symmetric KL divergence $s'$, which corresponds to the minimum change in distribution that is to be detected in a streaming data setting. This minimum symmetric KL divergence can be used to set values of window size $w^*$ and drift term $v^*$.  The optimal window size $w^* $ can be found by minimizing an expression. This expression is discussed in more detail in Remark \ref{remark: min w}. Despite this expression being convex with respect to $w$, a closed-form expression of $w^*$ is difficult to obtain. This optimal window size $w^*$ can be solved numerically.
When a change point is detected, the previous post-change estimate $\hat{\theta}_t$ is used as the pre-change distribution $\theta_0$ for detecting the subsequent change point.


\begin{algorithm}[H]
\caption{DAS-CUSUM for multiple change point detection}
\begin{algorithmic}
\label{Alg: 1}
\State \textbf{Inputs:}  Sequence: $X$, Threshold $b$,  Target ARL: $\gamma$, Min sym div: $s'$, Pre-change dist :$\theta_0{(\mu_0,\sigma_0^2)}$
\State \textbf{Output} CpList : List containing change points
\State Choose window size
\begin{align*}
w^* &=  \mathop{\text{arg\,min}}_w  \frac{\log \gamma }{- 1 + (1 + ws'^2)^{\frac{1}{2}} + \log\left(1 - \frac{\left(-1 + (1+ws'^2)^{\frac{1}{2}}\right)^2 }{ws'^2}\right)} + w \\
\deln^* &= -\frac{1}{s'} + \left(\frac{1}{s'^2} + w^*\right)^{1/2}, \quad v* =  \frac{-\log(1 - \frac{{\deln^*}^2}{w})}{\deln^*}
\end{align*}
\State \textbf{}
    \For{$t = 1 \text{ to length}(X)]$}
    \State $\hat{\mu}_t = \sum_{i=t+1}^{t+w} x_i,\quad \hat{\sigma}_t^2 = \sum_{i=t+1}^{t+w} \frac{1}{w} ( x_i - \hat{\mu}_t  )^2, \quad \hat{\theta}_t = \{\hat{\mu}_t, \hat{\sigma}^2_t\}$
    \State Compute CUSUM recursion \[S_t = ( S_{t-1} )^+ + \log{\frac{f_{\hat{\theta}_t}(x_i)}{f_{\theta_0}(x_i)}} +  D_{\rm KL}(f_{\theta_0}(x)  , f_{\hat{\theta}_t}(x) ) - v^*\]
        \If{$S_t > b$}
        \State Add $t$ to CpList
        \State $\mu_0 = \hat{\mu}_t, \quad \sigma_0^2 = \hat{\sigma}^2_t$
        \EndIf
    \EndFor
\end{algorithmic}
\end{algorithm}

\section{Theoretical results: EDD versus ARL }
\label{sec: theoratical}
When our detection delay takes $T$ samples to detect a change, the average run length (ARL) is the expected value of $T$ under the pre-change distribution $\theta_0$ such that a false change is detected. Expected detection delay (EDD) is the expected value of $T$ under the post-change distribution $\theta_1$ such that a true change point is detected. 
Our first result relates  DAS-CUSUM's average run length with its expected detection delay. Similar analysis techniques have been done in \cite{xie2020sequential,xie2022window}.

\begin{theorem}{\label{Theorem:1}Let $f_{\theta_0}(x) $ and $f_{\theta_1}(x)$ be  the Gaussian probability density functions of $x$ under the pre-change distribution $\theta_0$ and post-change distribution $\theta_1$ which is unknown and estimated using a window of size $w$. $D_{\rm KL}(f_{\theta_1}(x) , f_{\theta_0}(x) )$ is the Kullback-Leibler divergence between these distributions and $\deln > 0$ .  For the proposed statistic, the asymptotic(as $w$ becomes large) expected detection delay (EDD) for a change from $ x \sim \mathcal{N}(\theta_0 )$ to  $ x \sim \mathcal{N}(\theta_1 ) $  at an average run length (ARL) $\gamma$, with $\gamma \rightarrow \infty$ is given by:
\begin{align*}
     \mathbb{E}_{\theta_1}[T] &=  \frac{\log \gamma +o(1)}{\deln \left( D_{\rm KL}(\theta_1,\theta_0) +  D_{\rm KL}(\theta_0,\theta_1)\right) + \log(1 - \frac{{\deln}^2}{w})} + w   .
\end{align*}
}\end{theorem}

\begin{corollary}{ 
\label{Corollary: opt_delta}
The value of $\deln$ that minimizes the expected detection delay for a given average run length in Theorem \ref{Theorem:1} is given by:
\begin{align*}{
    \deln^* = -\frac{1}{(D_{KL} (\theta_0, \theta_1) + D_{\rm KL} (\theta_1, \theta_0) } + \left(\frac{1}{\left((D_{\rm KL} (\theta_0, \theta_1) + D_{\rm KL} (\theta_1, \theta_0)\right)^2} + w \right)^{1/2} .
}\end{align*}
}\end{corollary}

\begin{remark}
\label{remark: optimal}
\normalfont The value of $\deln^*$ from Corollary \ref{Corollary: opt_delta} can be used in the result of Theorem \ref{Theorem:1} to obtain the minimum EDD for a given ARL.
\end{remark}
\begin{corollary}
\label{Corollary: opt_drift}
{The optimal drift term $v^*$ which minimizes the expected detection delay for any ARL is given by
\[
     v^* = \frac{-\log(1 - \frac{\deln^*{^2}}{w})}{\deln^*}.
\]
}\end{corollary}

\begin{figure}
    \centering
    \includegraphics[width = .45\textwidth]{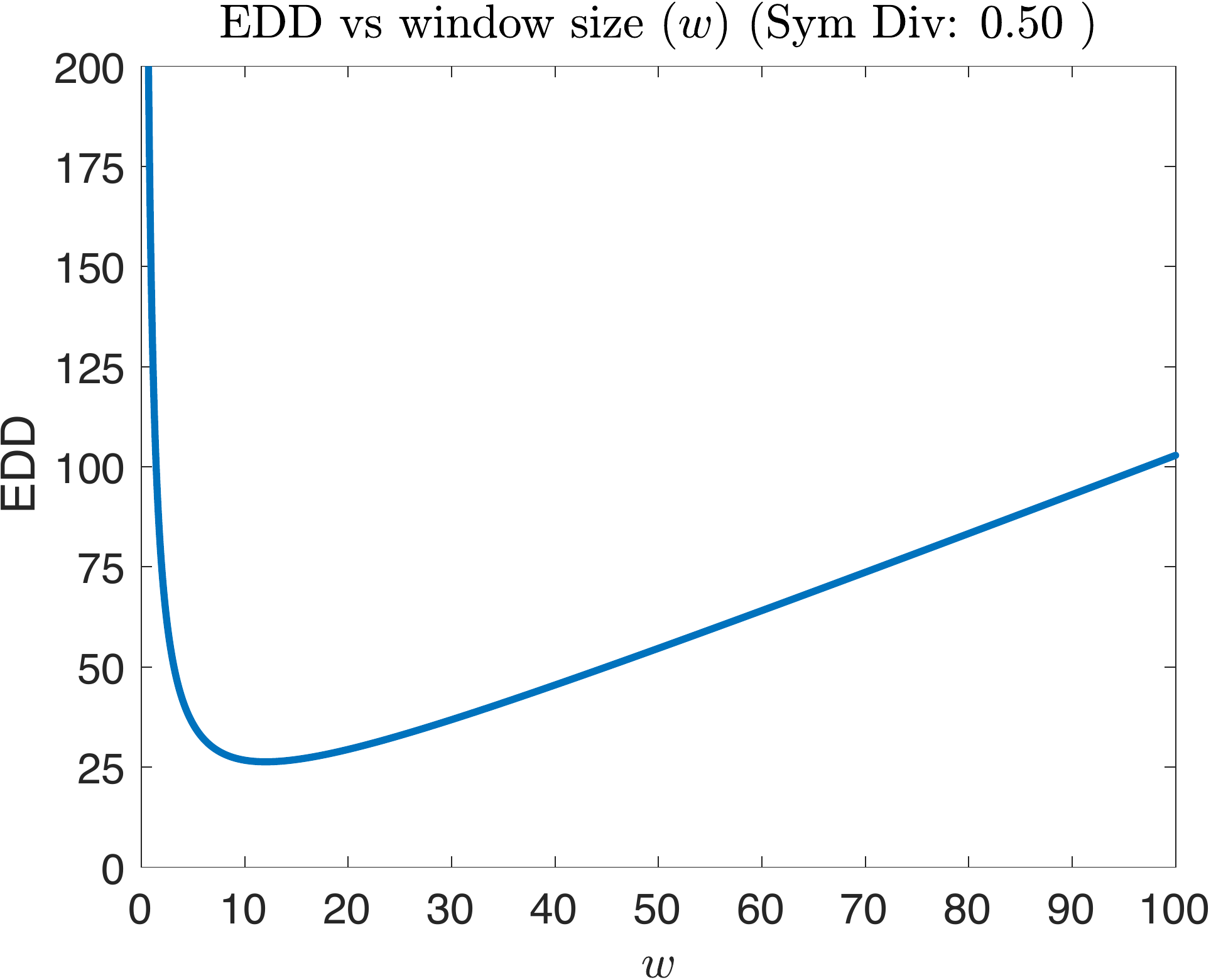}
    \caption{Expected detection delay (EDD) versus window size for a change with symmetric KL divergence of 0.5 at an ARL value of 5,000. This figure shows that the EDD is a convex function of $w$ which can be minimized to obtain the optimal window $w^*$.}
    \label{fig: edd versus w (sym :0.5) }
\end{figure}

\begin{remark}
\label{remark: min w}
\normalfont The expression in Theorem \ref{Theorem:1} can be minimized with respect to $w$ (at a provided value of average run length and symmetric KL divergence) to find the optimal window size $w^*$. A closed-form expression for $w^*$ cannot be obtained, but $w^*$ can be solved numerically. Figure \ref{fig: edd versus w (sym :0.5) } shows how EDD relates to window size $w$. The curve has a minimum point which corresponds to a window size of $w=11$.  When this solution is too small, the results in Theorem \ref{Theorem:1}  do not hold, which assume $w$ to be large (so that post-change estimates converge to true post-change distribution). More details on this can be found in Section \ref{sec: opt win length}. Additionally, the window size should be large enough such that $\deln^* < w$ for the logarithmic term in Theorem \ref{Theorem:1} to be real.
\end{remark}

\subsection{Comparison to CUSUM results}
\label{sec: comparison to CUSUM results}
The relationship between ARL ($\gamma$) and EDD ($\mathbb{E}_{\theta_1}(T)$ for CUSUM \cite{lorden1971procedures} is shown below:
\begin{equation*}
    \mathbb{E}_{\theta_1}(T) = \frac{\log\gamma(1+o(1))}{\mathbb{E}_{\theta_1}\left[\log\frac{f_{\theta_1}(x)}{f_{\theta_0}( x ) }\right]} .
\end{equation*}

For the proposed statistic, it can be seen in Theorem~\ref{Theorem:1} that the expected detection delay at a set ARL value would be similar for a change from $\theta_0$ to $\theta_1$ and a change from $\theta_1$ to $\theta_0$. This is not true for CUSUM, where the detection delay for a change from $\theta_0$ to $\theta_1$ will not be equal to a change from $\theta_1$ to $\theta_0$.

The expression in Theorem 1 also has an additional $w$ term, which takes into account the time delay for obtaining the window to estimate post-change parameters, but this is a consequence of the post-change distribution is unknown.

\subsection{Sketch of the Proof}

The increment of the CUSUM statistic in~\eqref{eq: CUSUM} consists of a log-likelihood ratio which has a negative expectation under the pre-change distribution $\theta_0$. The proposed increment statistic for DAS-CUSUM in~\eqref{eq: DAS-CUSUM update} has a negative drift under the post-change distribution but is not a log-likelihood ratio. 
One way to find the optimal value $v$ in our proposed update statistic is to convert it into a valid log-likelihood ratio. Once this is done, ARL and EDD results from CUSUM can be used for our proposed statistic. This expression would consist of the negative drift term $v$, which could be minimized to find the optimal value for $v$.
It can been seen in \cite{lorden1971procedures} that for a detection threshold $b$ , the CUSUM procedure has the following average run length:
\begin{equation}
    \mathbb{E}_{0}[T] = \frac{e^{b}( 1 + o(1))}{K} ,
    \label{eq : ARL CUSUM}
\end{equation}
where $K$ is a constant. 
For CUSUM, the expected detection delay is related to the detection threshold $b$ by
\begin{equation}
\label{eq : CUSUM EDD}
    \mathbb{E}_{\theta_{1}}(T) = \frac{ b + o(1)}{\mathbb{E}_{\theta_1}\left[\log\frac{f_{\theta_1}(x)}{f_{\theta_0}( x ) }\right]}.
\end{equation}
Using the tools proposed in \cite{xie2018first}, an equivalence term $\deln$ can be introduced to our incremental statistic, which satisfies the equation
\begin{equation}
\label{eq : martingale}
    \mathbb{E}_{\theta_0}[\exp({\deln {s}_t})] = 1.
\end{equation}
When~\eqref{eq : martingale}
is satisfied, $\deln s_t$ is a martingale and can be considered to be the log-likelihood ratio between distributions $\Tilde{f}_{\theta_1} = \exp[{\deln s_t}]f_{\theta_0}$ and $f_{\theta_0}$ which then allows us to use~\eqref{eq : ARL CUSUM} to obtain the ARL performance for DAS-CUSUM. The threshold $b$ 
can be expressed in terms of the average run length ($\gamma$) 
\begin{equation}
\label{eq: thresh arl DAS-CUSUM}
    b = \frac{\log \gamma + o(1)}{\deln}.
\end{equation} 
This expression is obtained through~\eqref{eq : ARL CUSUM}  where the constant $K$ is absorbed within $o(1)$ and the introduced scaling factor $\delta_0$ for the incremental statistic is appropriately scaled.
Similarly, $\delta_{1}$ can be introduced such that $\delta_1 s_t$ is the log-likelihood ratio between $f_{\theta_1}$ and $\Tilde{f}_{\theta_0} = \exp[-{\delta_1 s_t}]f_{\theta_1}$. Thus~\eqref{eq : CUSUM EDD} can be used to relate change between $f_{\theta_1}$, where the $\delta_1$ term is observed in $o(1)$ as shown below:

\begin{equation}
   \mathbb{E}_{\theta_{1}}(T) = \frac{ b + o(1)}{\mathbb{E}_{\theta_1}\left[s_t\right]} .
\end{equation}
Substituting~\eqref{eq: thresh arl DAS-CUSUM} in the above  equation, we obtain
\begin{equation}
\label{eq: EDD ARL DAS-CUSUM}
   \mathbb{E}_{\theta_{1}}(T) = \frac{ \log \gamma + o(1)}{\deln \mathbb{E}_{\theta_1}\left[s_t\right]}.
\end{equation}
Substituting~\eqref{eq: expect under postchange} yields
\begin{equation}
\label{eq: EDD ARL DAS-CUSUM div}
   \mathbb{E}_{\theta_{1}}(T) = \frac{ \log \gamma + o(1)}{\deln \left( D_{\rm KL}(\theta_0 , \theta_1) + D_{\rm KL}(\theta_1, \theta_0) -v \right) }.
\end{equation}

Our expression above assumes that our statistic is converted to a log-likelihood ratio by satisfying the martingale property in~\eqref{eq : martingale}. Lemma \ref{lemma: v equiv} satisfies this requirement by finding an expression that relates the drift value $v$ with the equivalence factor $\deln$

\begin{lemma} {
As $w \xrightarrow{}  \infty, \mathbb{E}_{\theta_0}[\exp({\deln {s}_t})] = 1$  when
\begin{equation}
   v =  \frac{-\log(1 - \frac{\deln{^2}}{w})}{\deln} 
   \label{eq: v value}.
\end{equation}
\label{lemma: v equiv}
}\end{lemma}
The value for $v$, for which \eqref{eq : martingale} is satisfied, can be substituted. As $w$ samples are needed to estimate the post-change distribution $\hat{\theta_t}$, the detection delay would be

\begin{equation}
\label{eq: before optim EDD vs ARL DAS-CUSUM }
   \mathbb{E}_{\theta_{1}}(T) = \frac{ \log \gamma + o(1)}{\deln \left( D_{KL}(\theta_0 , \theta_1) + D_{KL}(\theta_1, \theta_0) \right) + \log(1 - \frac{\deln{^2}}{w})  } + w.
\end{equation}
This expression can be minimized with respect to $\deln$ by taking the derivative and equating to 0. The resulting optimal value of $\deln^*$. is given below:
\begin{equation}
    \deln^* = -\frac{1}{(D_{KL} (\theta_0, \theta_1) + D_{KL} (\theta_1, \theta_0) } + \left(\frac{1}{\left((D_{KL} (\theta_0, \theta_1) + D_{KL} (\theta_1, \theta_0)\right)^2} + n\right)^{1/2}
    \label{opt: deln}.
\end{equation}
Using this optimal value of $\deln$ in~\eqref{eq: before optim EDD vs ARL DAS-CUSUM } and~\eqref{eq: v value} leads to the results of Theorem \ref{Theorem:1} and Corollary \ref{Corollary: opt_drift}.

\subsubsection{Sketch of Proof for Lemma 1}

The left side of~\eqref{eq : martingale} can be written as shown below by substituting the proposed update statistic from~\eqref{eq: DAS-CUSUM update}:
\begin{equation*}
    \mathbb{E}_{\theta_0}\left[\exp({\deln \Tilde{s}_t})\right] =  \mathbb{E}_{\theta_0}\left[\exp\left(\deln \left( - \frac{(x_t - \mut)^2}{2 \sigt^2}+\frac{(x_t-\mu_0)^2}{2\sigma_0^2} + \frac{\sigma_0^2 + (\mu_0 - \mut)^2}{2\sigt^2} - \frac{1}{2} -v \right)\right)\right].
\end{equation*}
Since a future window ($x_{t+1}....x_{t+w}$) is used to estimate $\mut$ and $\sigt$, these estimates are independent from $x_t$.  These estimates can be treated as constants while introducing a conditional expectation through the tower rule. The equation above can be written as

\begin{align}
   &\mathbb{E}_{x_{t+1,...,t+w}\sim \theta_0}\left[\exp\left(\deln\left(\frac{\sigma_0^2 + (\mu_0 - \mut)^2}{2\sigt^2} - \frac{1}{2} -v \right) \right) \mathbb{E}_{x_t \sim \theta_0}\left[  r(x_t)  \Big| \mut,\sigt\right]\right] \nonumber \\
   &=\exp(\deln(-\frac{1}{2} -v))\mathbb{E}_{\mathop{x}_{t+1,..,t+w}\sim \theta_0}\left[\exp\left(\deln\left(\frac{\sigma_0^2 + (\mu_0 - \hat{\mu})^2}{2\sigt^2} \right) \right) \mathbb{E}_{x_t \sim \theta_0}\left[ r\left(x_t\right) \Big| \mut,\sigt\right]\right]
   \label{eq: Outer expec},
\end{align}
where
\begin{equation*}
    r \left( x_t \right) = \exp \left( \deln \left( - \frac{(x_t - \mut)^2}{2 \sigt^2}+\frac{(x_t-\mu_0)^2}{2\sigma_0^2} \right) \right) .
\end{equation*}



Further details for these calculations can be found in the Appendix.

\section{Simulations}
\label{sec: simulations}

\subsection{ARL and EDD}

\begin{figure}%
    \centering
    \subfigure[ ]
     {  \centering
        \label{fig: DAS-CUSUM_vs_CUSUM_w10}%
        \includegraphics[height=.42\textwidth]{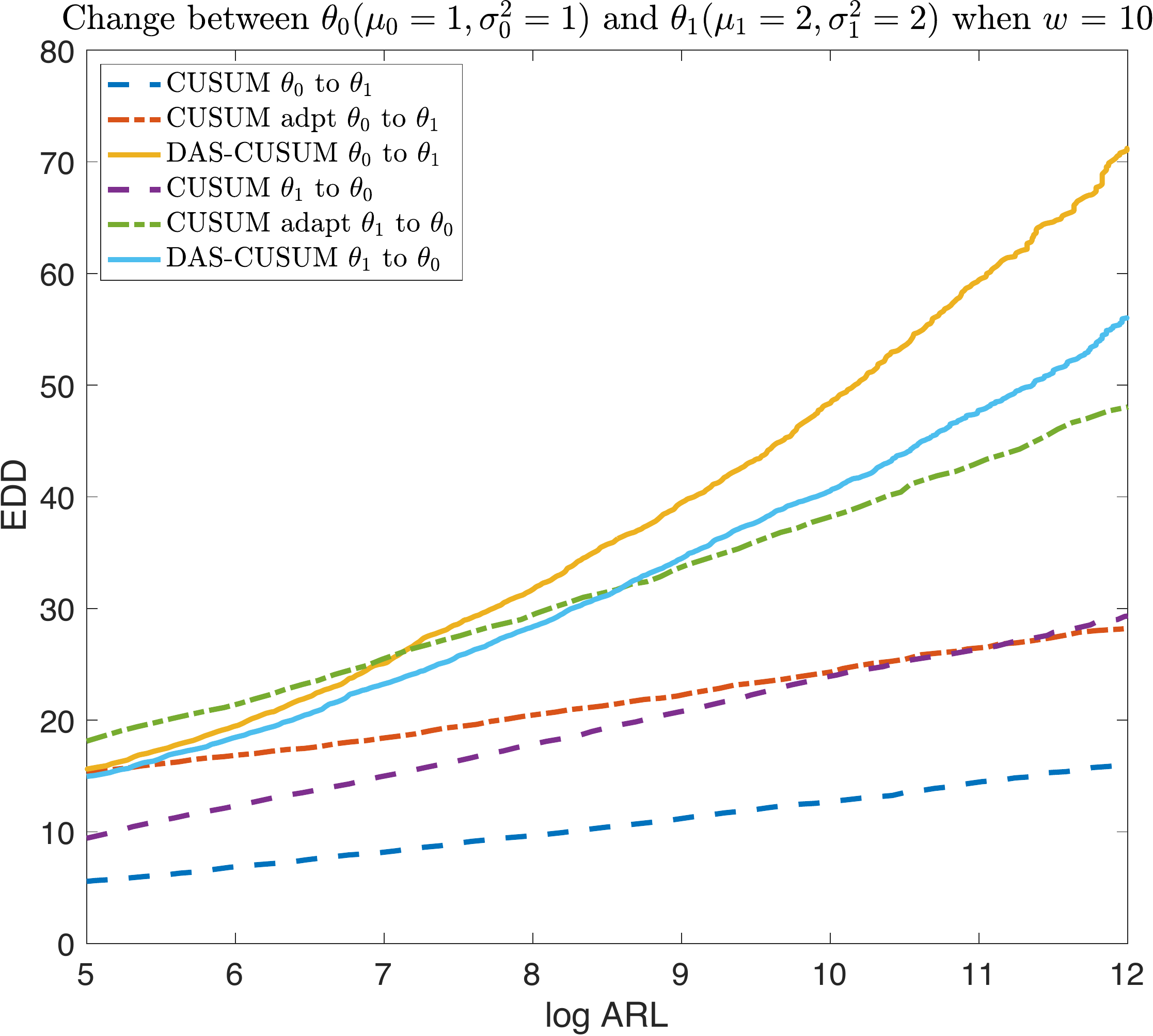}
    }
    \subfigure[ ]
     {  \centering
        \label{fig: DAS-CUSUM_vs_CUSUM_w40}%
        \includegraphics[height=.42\textwidth]{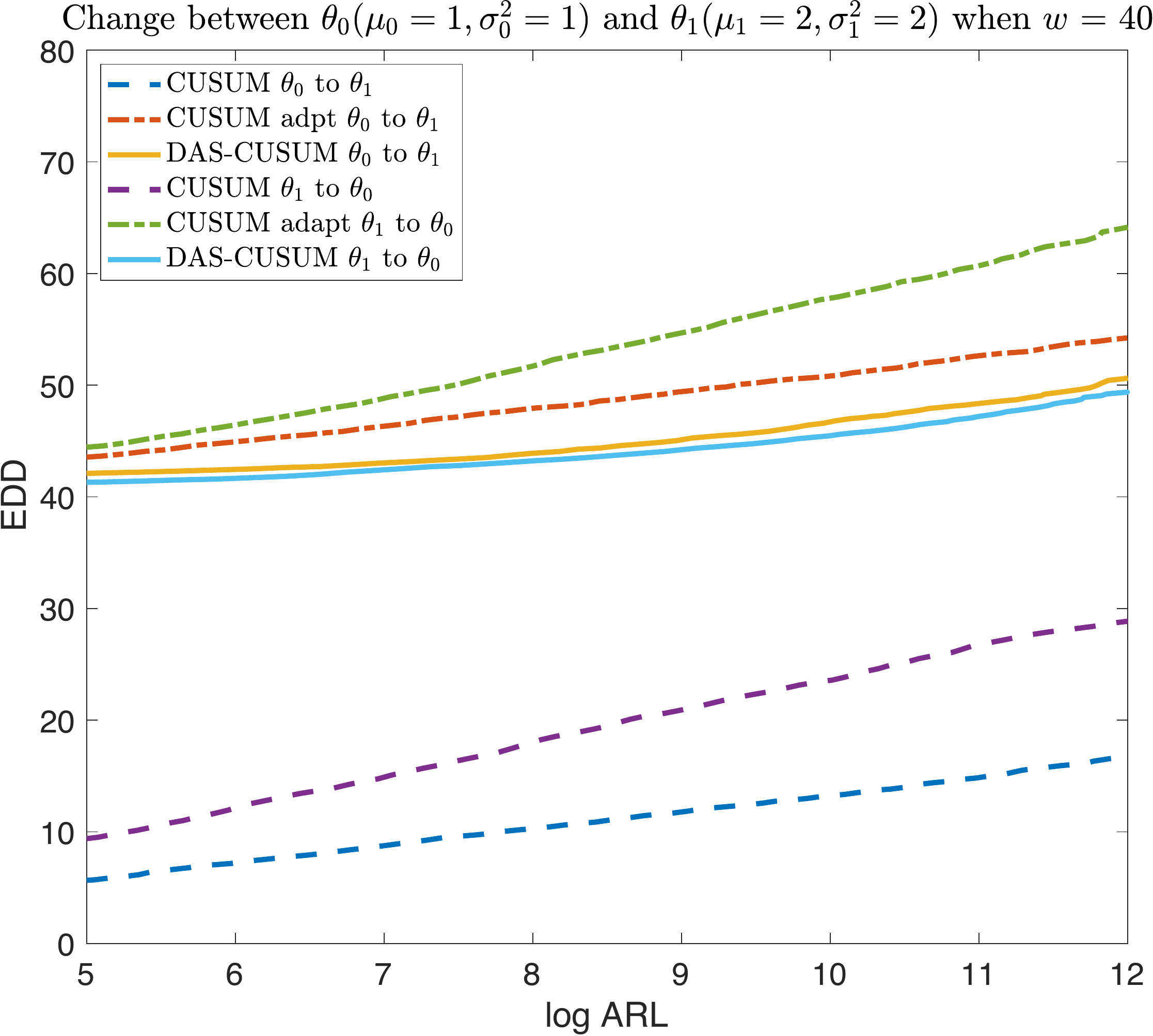}
    }   \caption{EDD versus ARL performance comparison for DAS-CUSUM and CUSUM for changes between $\theta_0 (\mu_0 =1,\sigma_0^2 =1 )$ and $\theta_1 (\mu_1 =2, \sigma_1^2 =2 )$ which corresponds to a symmetric KL divergence of 1. Figure \ref{fig: DAS-CUSUM_vs_CUSUM_w10} shows the relationship when a  window size of 10  is used for the post-change estimate. while Figure \ref{fig: DAS-CUSUM_vs_CUSUM_w40} shows the case when the window size is 40. Notice the similar performance for DAS-CUSUM for changes from $\theta_0$ to $\theta_1$ and $\theta_1$ to $\theta_0$. This similarity increases with window size $w$.}
   \label{fig: symmetric vs. CUSUM}
\end{figure}

As discussed in Section \ref{sec: proposed statistic}, the DAS-CUSUM change point detection procedure is designed to have a symmetric change statistic.  Due to this symmetric property, DAS-CUSUM should have similar ARL versus EDD performance for changes from the distribution $\theta_0$ to $\theta_1$ and from $\theta_1$ to $\theta_0$. This symmetry is studied in ARL versus EDD plots in Figure \ref{fig: symmetric vs. CUSUM}. This figure also contains plots for CUSUM and an adaptive version of CUSUM where a future window of size $w$ is used to estimate the post-change parameters. CUSUM curves for changes from $\theta_0 (\mu_0 = 1, \sigma_0^2=1)$ to $\theta_1 (\mu_1 = 2, \sigma_1^2=2)$ and $\theta_1$ to $\theta_0$ are far away from one another, while DAS-CUSUM curves are closer to each other. These DAS-CUSUM curves become closer when the post-change estimates become more accurate with an increasing window size, as shown in Figure \ref{fig: DAS-CUSUM_vs_CUSUM_w40}. These results are is in line with Section \ref{sec: comparison to CUSUM results}, which compares the results of DAS-CUSUM in Theorem \ref{Theorem:1} with corresponding results for CUSUM. Specifically, EDD at a given ARL is the same for a change from $\theta_0$ to $\theta_1$ and vice versa when the window length $w$ becomes asymptotically large.

\begin{figure}[t]
    \centering
    \includegraphics[width = .6\textwidth]{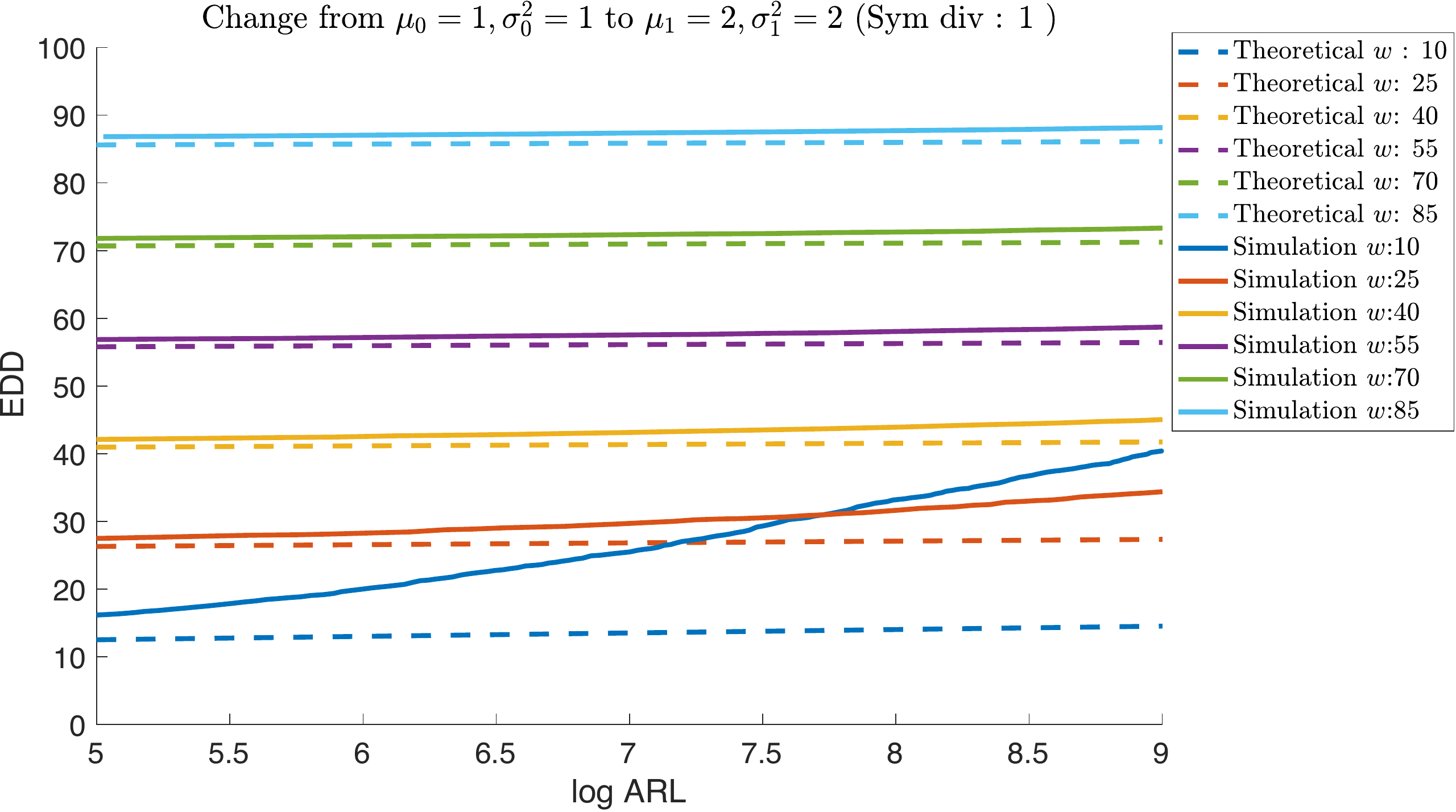}
    \caption{Comparison between theoretical and simulated DAS-CUSUM results for different post-change estimation window sizes $w$. The change in this example has a symmetric KL divergence of 1.}
    \label{fig: theoretical vs. simulation}
\end{figure}

Now we validate the accuracy of theoretical approximation by comparing it against simulation results. Figure \ref{fig: theoretical vs. simulation} shows DAS-CUSUM  plots for EDD versus ARL at different window lengths ($w$ to estimate post-change distribution).  For each window length, plots for the theoretical relationship (from Theorem \ref{Theorem:1}) are compared to simulated plots. For a small window size ($w=10$), the theoretical and simulated results grow apart as ARL increases. The difference between the theoretical and simulated plots decreases as the window size increases. This is expected as the results in Theorem \ref{Theorem:1} hold when $w$ grows asymptotically. When $w=120$, the difference between theoretical and simulated EDD is approximately 1 sample for the shown ARL range.

\subsection{Optimal window length}
\label{sec: opt win length}
\begin{figure}[ht]%
    \centering
    \subfigure[]
     {  \centering
        \label{fig: edd vs w (sym :0.1)}
        \includegraphics[width = .45\textwidth]{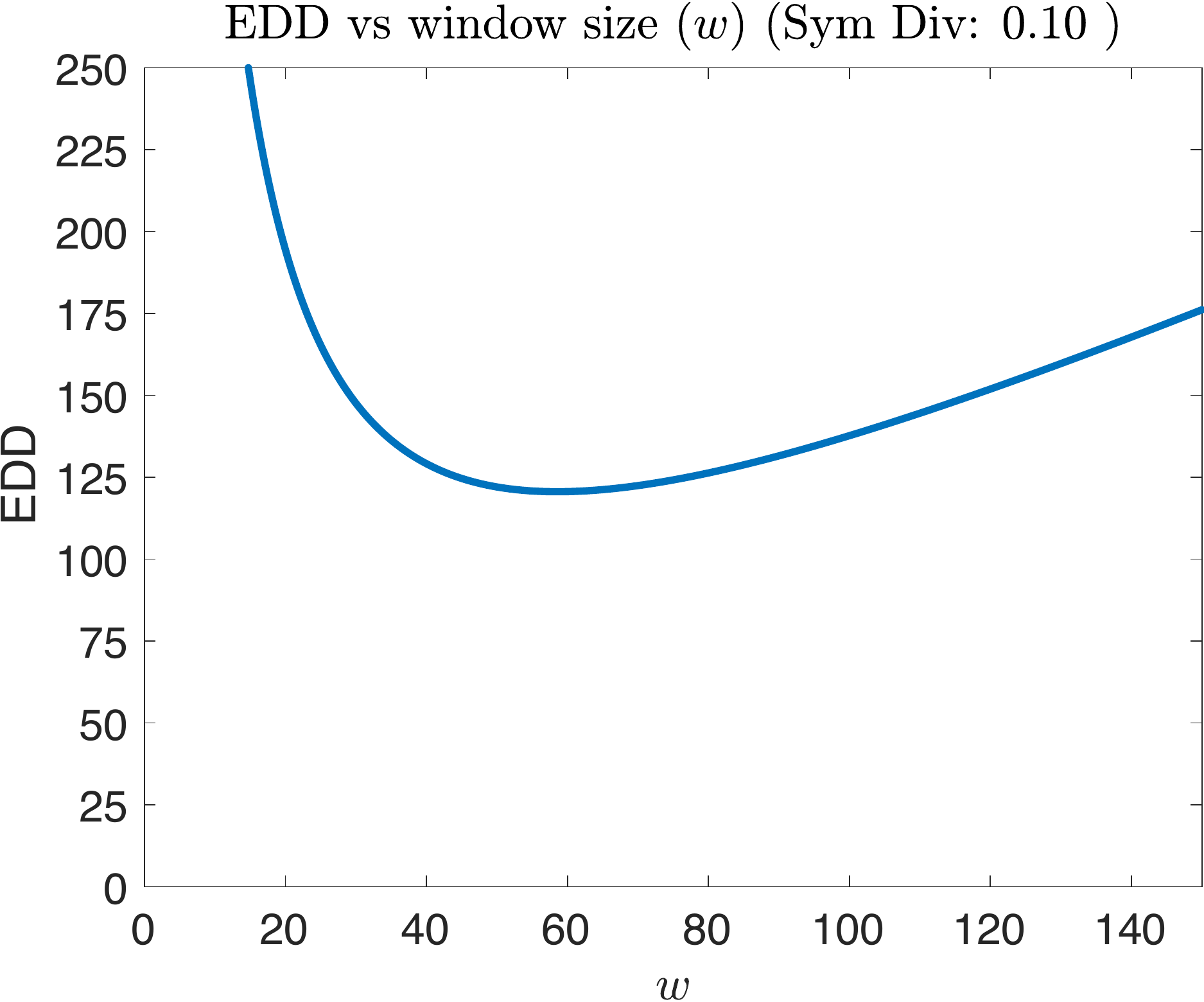}
    }
    \subfigure[ ]
     {  \centering
        \label{fig: edd vs w (sym :2)}
        \includegraphics[width = .45 \textwidth ]{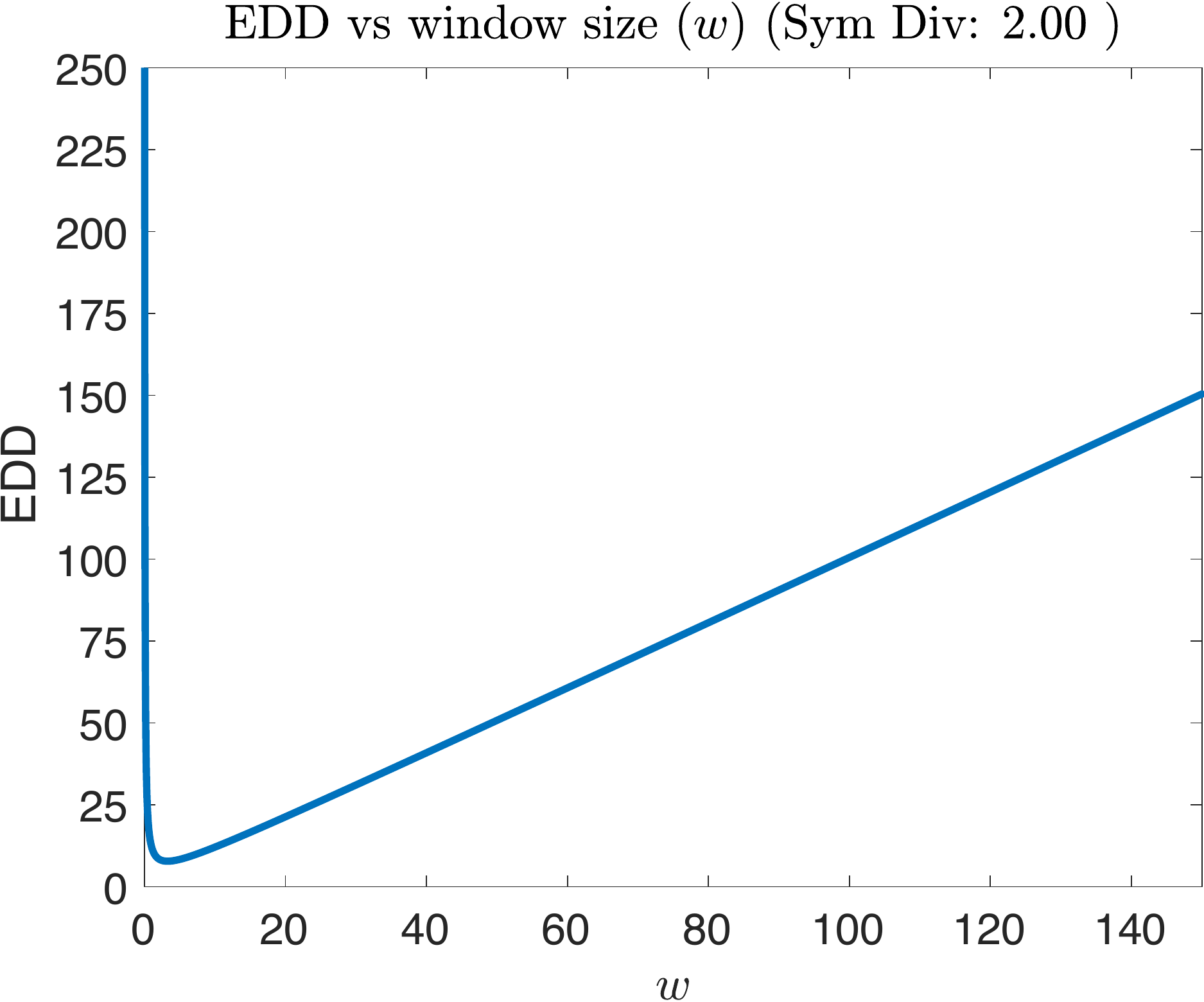}
    }   \caption{ Relation between EDD and window size for changes with different symmetric KL divergence. The ARL has been set to 5,000 in both figures. The optimal window size corresponds to the minimum EDD values. Figure \ref{fig: edd vs w (sym :0.1)} shows the relationship for a change with symmetric KL divergence of 1 while \ref{fig: edd vs w (sym :2)} shows the relationship for a symmetric KL divergence of 2}.
   \label{fig: edd vs w comparison}
\end{figure}

DAS-CUSUM results that relate EDD with ARL in Theorem \ref{Theorem:1}  depend on the estimation window size $w$ (at provided values of ARL and symmetric KL divergence). 
This equation can be minimized for $w$ to find the optimal window length ($w^*$). Unfortunately, there is no closed-form expression for this optimal value. Nevertheless, this equation can be minimized numerically to obtain $w^*$. Figure \ref{fig: edd vs w comparison} shows this relationship at an ARL of 5,000 for changes with two different symmetric KL divergence values.

Figure \ref{fig: edd vs w comparison} shows this relationship for a smaller change in distribution (a symmetric diverge of 0.11), while Figure \ref{fig: edd vs w (sym :2)} shows this relationship for a larger change (a symmetric KL divergence of 2). Intuitively, a larger change (with a larger symmetric KL divergence) would be easier to detect, requiring a shorter window length as compared to a smaller change (with a smaller symmetric KL divergence). However, for larger changes, the window size corresponding to the minimum EDD value could be too small, as seen in Figure \ref{fig: edd vs w (sym :2)} where this window is of size 4. The theoretical results start to match simulated results at a window size of about 30 while results at a window size of 10 divergences. For this reason, when the optimal window size ($w^*)$ is below 20, a rule should be in place for a minimum window size. 

\begin{figure}[H]%
    \centering
    \subfigure[]
     {  \centering
        \label{fig: opt_w_1dot3}
        \includegraphics[width = .48\textwidth]{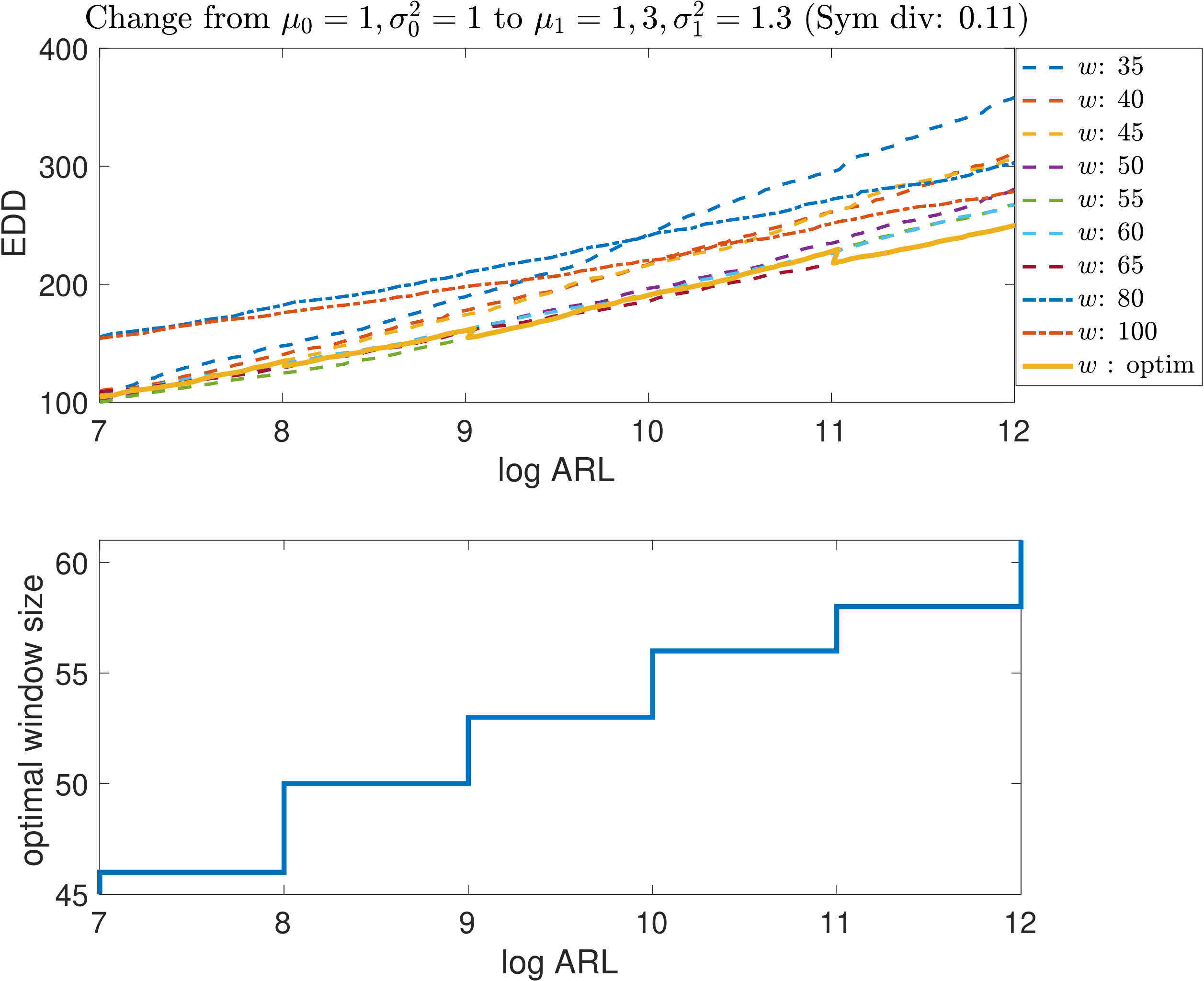}
    }
    \subfigure[ ]
     {  \centering
        \label{fig: opt_w_1dot2}
        \includegraphics[width = .48\textwidth]{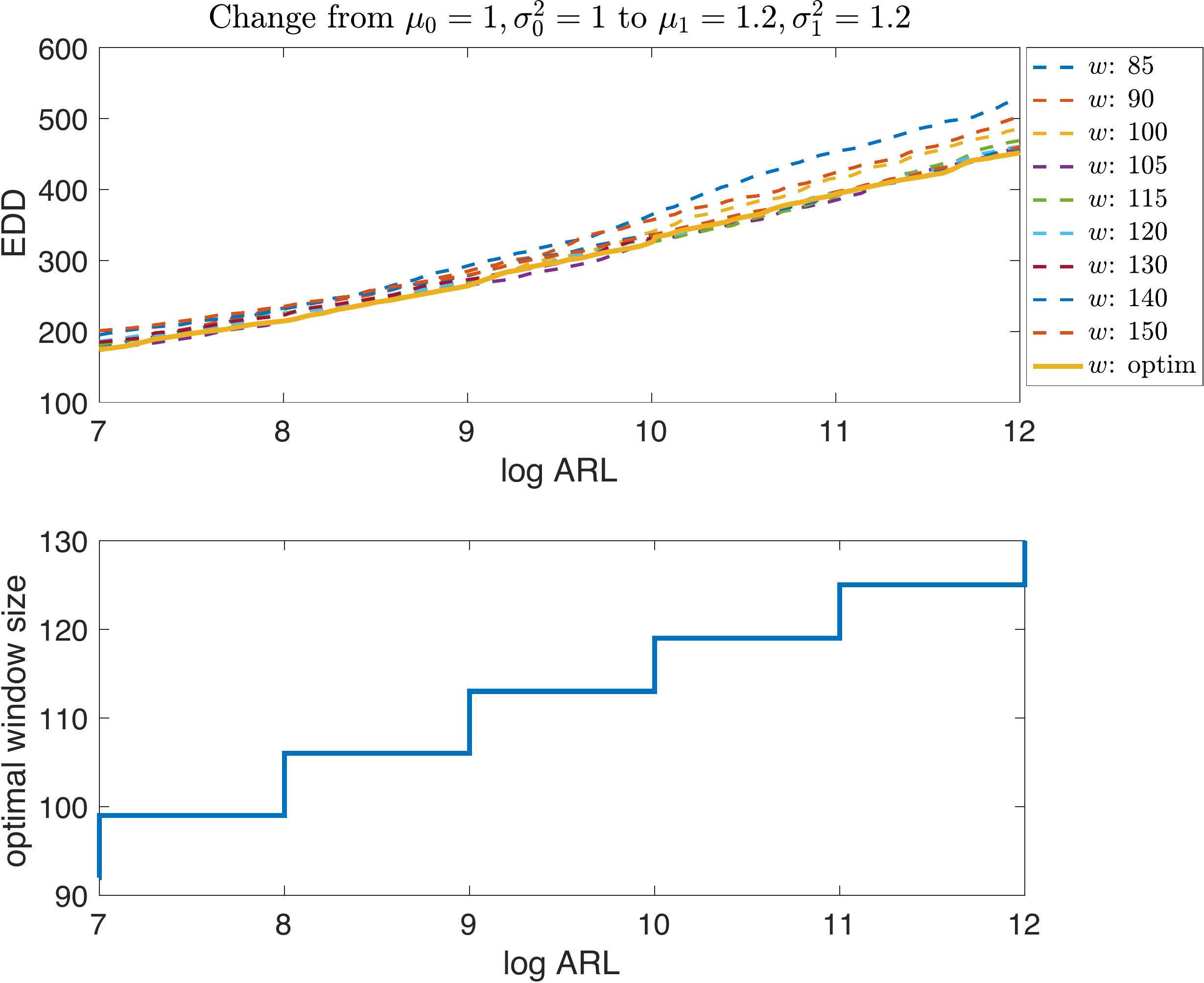}
    }   \caption{ ARL versus EDD performance for different window length ($w$) sizes. Figures \ref{fig: opt_w_1dot3} shows plots for a change from $\theta_0(\mu_0 = 1,\sigma_0^2 =1)$ to $\theta_1(\mu_1 = 1.3,\sigma_1^2 =1.3)$ and \ref{fig: opt_w_1dot2} shows plots for changes from $\theta_0(\mu_0 = 1,\sigma_0^2 =1)$ to $\theta_1(\mu_1 = 1.2,\sigma_1^2 =1.2)$. Optimal window size ($w^*$) provides optimal performance as $w^*$ increases. }
   \label{fig: advantage DAS-CUSUM}
\end{figure}

Results that relate the optimal window length for different ARL values can be seen in Figure \ref{fig: edd vs w comparison}. The changes in this figure have small divergence values, which lead to $w^*$ that is greater than a size of 20. The curves for $w^*$ are in yellow and seem to provide better EDD versus ARL performance than most other window sizes. As the optimal window size, $w^*$ increases in Figure \ref{fig: opt_w_1dot2}, the corresponding ARL versus EDD curve often performs best (or close to best) when compared with other window sizes.

\subsection{Setting the detection threshold }

The table below compares the simulated and theoretical detection threshold ($b$) to achieve different ARL values. The theoretical relationship between ARL and the detection threshold is provided in~\eqref{eq: thresh arl DAS-CUSUM}. These experiments were done on a distribution change from $\theta_0 (\mu_0 = 1 , \sigma_0^2 =1 )$ to $\theta_1 (\mu_1 = 2 , \sigma_1^2 =2 )$ which corresponds to a symmetric KL divergence of 1. For ARL, false alarms occur when data points generated from pre-change distribution ($\theta_0$) is falsely detected as change points. Intuitively ARL values should depend only on the pre-change $\theta_0$  but the post-change distribution ($\theta_1$) is used to set the  $\deln^*$ value, which is used within the theoretical~\eqref{eq: thresh arl DAS-CUSUM} as well as for setting the drift term $v$ for the simulations. The difference between theoretical and simulated results is large for small values of post-change estimate window $w$, but these results become closer as this window size increases. This is expected as the relationship between the detection threshold, and average run length is obtained using~\eqref{eq : martingale}, which is satisfied asymptotically.
    \begin{table}[H]
        \centering
       \caption{Comparison between theoretical and simulated detection thresholds at different ARL values for a change with symmetric KL -divergence of 1} \begin{tabular}{lcccccccc}
            \toprule
             &  &$w = 10$  & $w=20 $  & $w=30$ & $w = 40$ & $w=50$ & $ w = 100$& $ w = 150$  \\
            \midrule
            ARL = 5,000 & Thr.  &  3.68 & 2.38 & 1.86 &1.57  & 1.37 & 0.94  & 0.75    \\
            & Sim. &  14.77  & 6.10 & 3.16 & 2.13  & 1.69 & 1.01  &  0.77 \\
              \hline
            ARL = 10,000 &  Thr.  &  3.98 & 2.57 & 2.02 &1.70  & 1.50 & 1.02  & 0.82   \\
            & Sim. &  18.16  & 7.91 & 4.13 & 2.70 & 2.11 & 1.26  &  0.96  \\
        \bottomrule
    \end{tabular}
  
  \label{Table: arl vs b}
\end{table}

\begin{figure}%
    \centering
    \subfigure[]
     {  \centering
        \label{fig: thresh vs w for ARL 5000}
        \includegraphics[width = .45\textwidth]{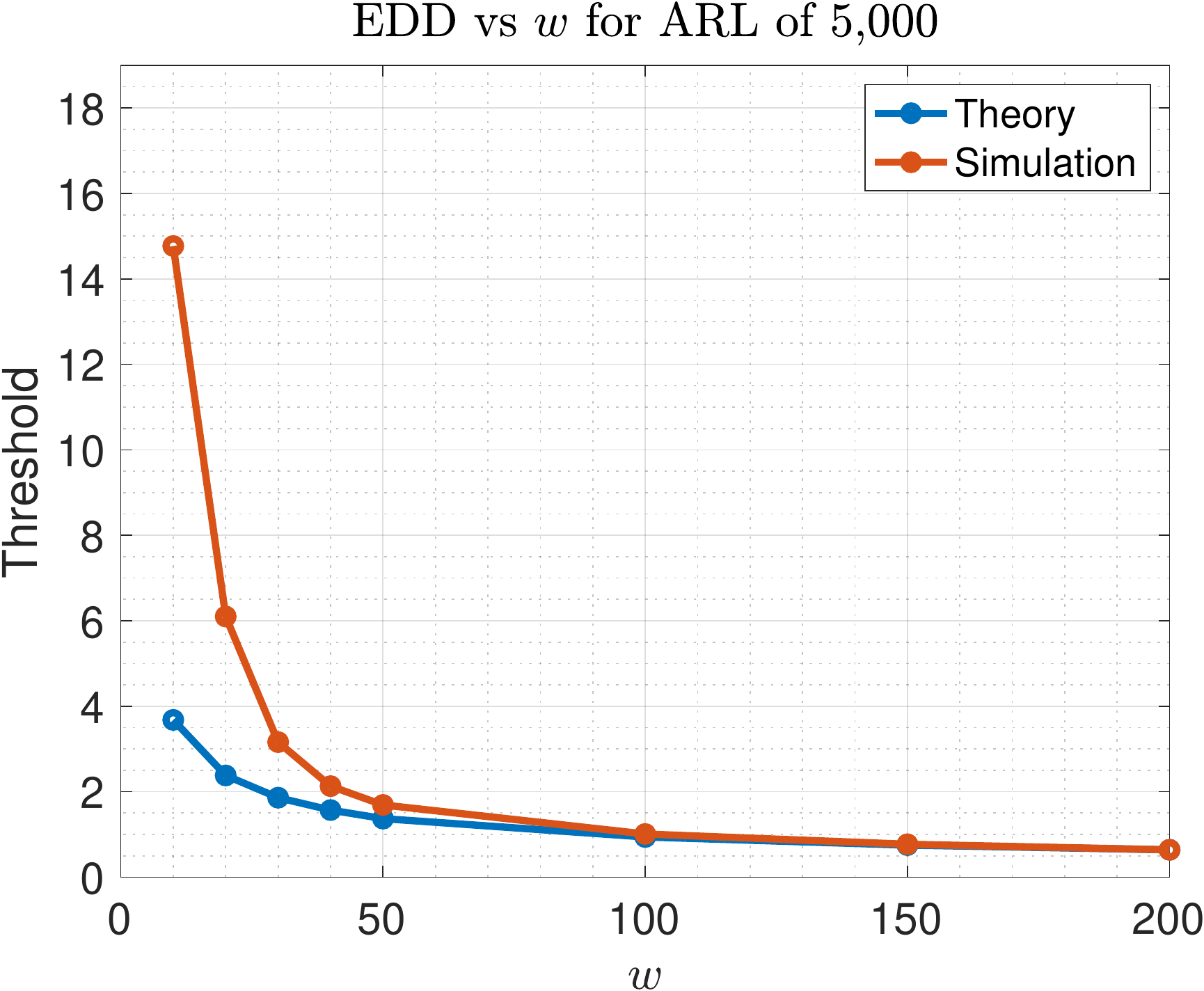}
    }
    \subfigure[ ]
     {  \centering
        \label{fig: thresh vs w for ARL 10000}
        \includegraphics[width = .45\textwidth]{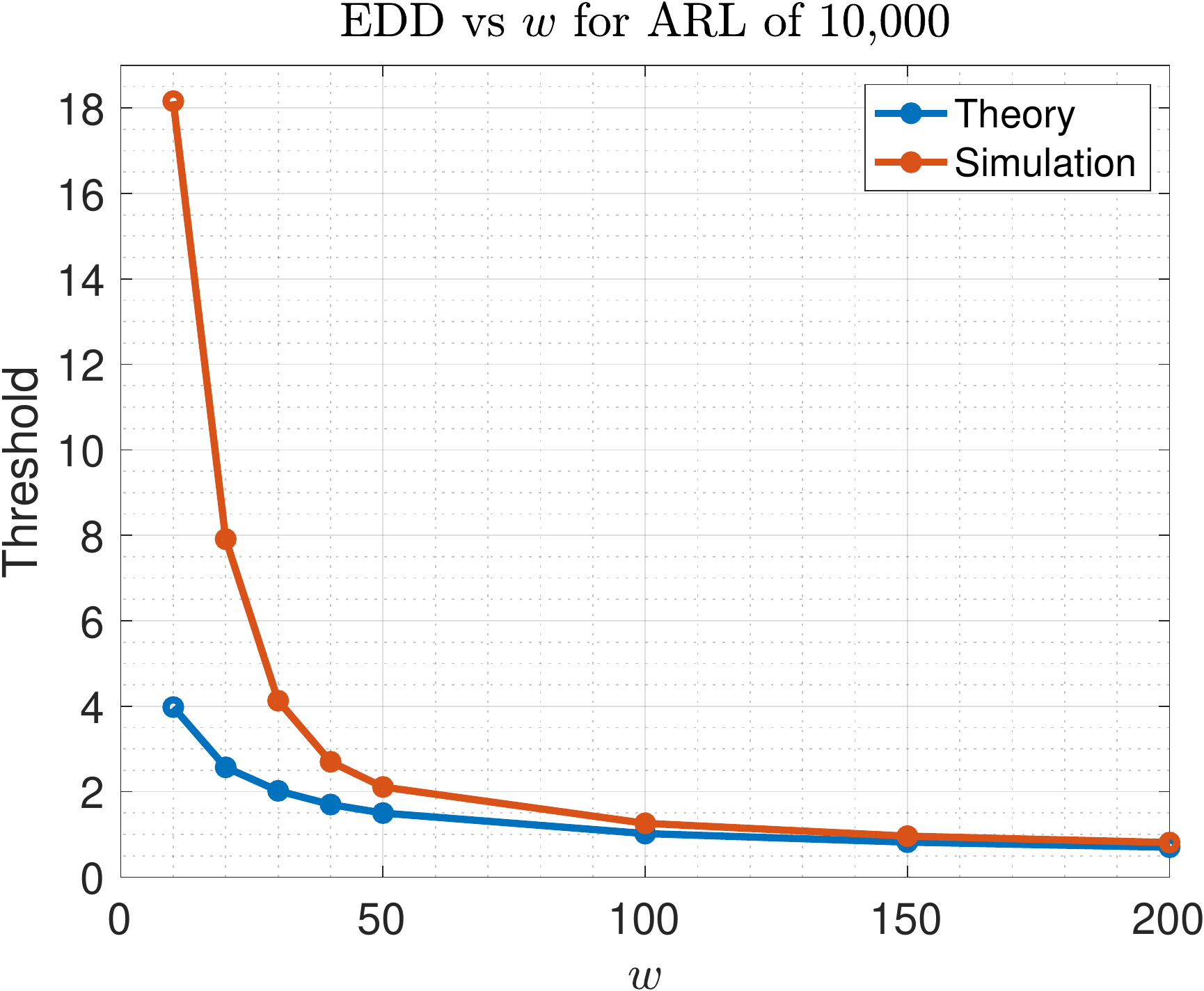}
    }   \caption{Plots for results in table \ref{Table: arl vs b}. Figure \ref{fig: thresh vs w for ARL 5000} shows the relationship for an ARL value of 5,000 while Figure \ref{fig: thresh vs w for ARL 10000} shows the relationship for an ARL of 10,000.. The gap between simulation and theoretical results gets small at a window value of about 30.}
   \label{fig: thresh vs w}
\end{figure}

\section{Real data}
\label{sec: real data}

\begin{figure}[t]
    \centering
    \includegraphics[width = 5 cm]{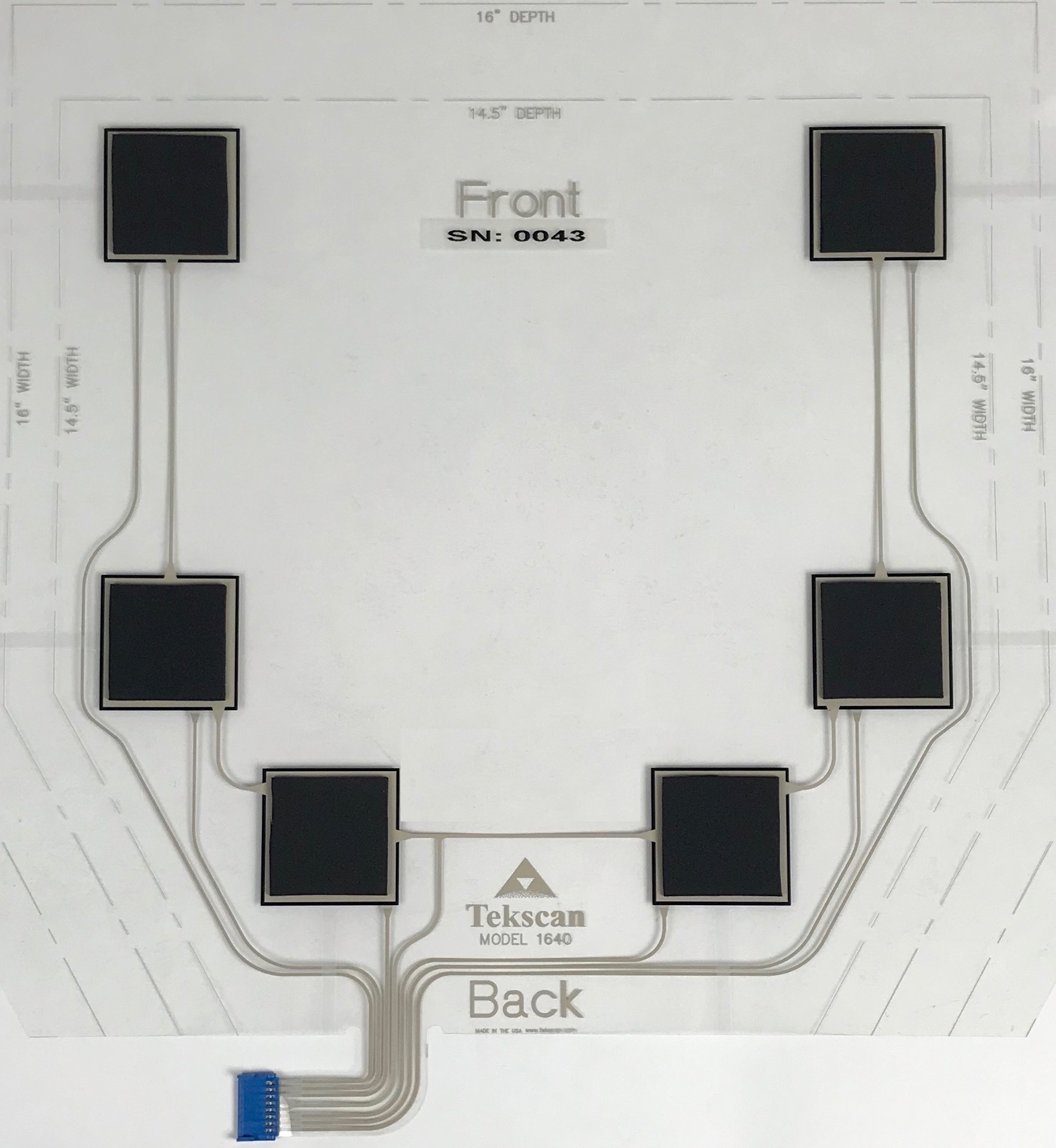}
    \caption{Sensor mat used for characterizing in-seat behaviour for wheelchair users. Sequential change point detection can be used to identify changes in wheelchair occupancy.}
    \label{fig: WiSAT}
\end{figure}

\begin{figure}%
    \centering
    \subfigure[]
     {  \centering
        \label{fig: sym_real_CUSUM}%
        \includegraphics[width = .45\textwidth]{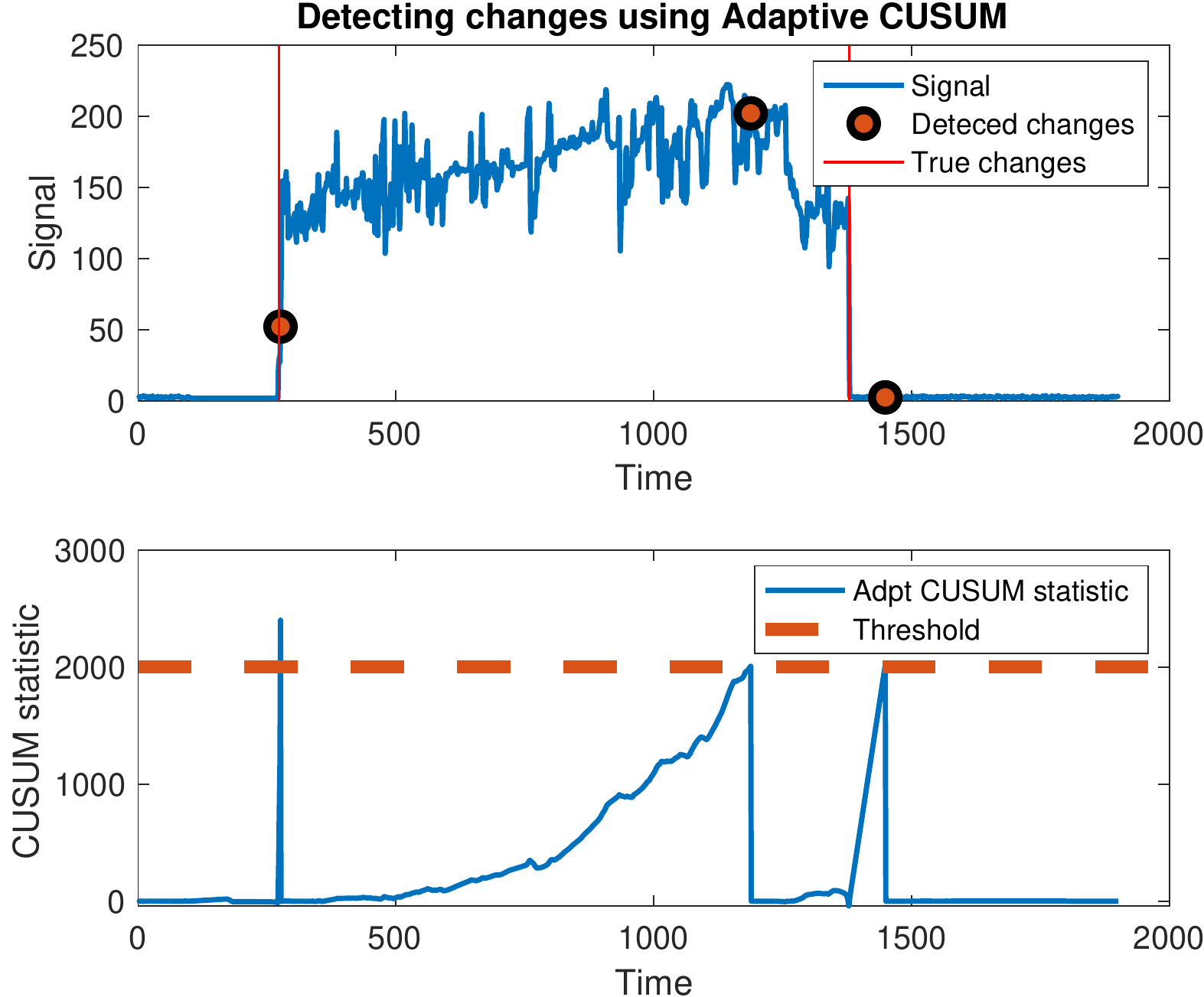}
        \label{fig: asymmetry wc}
    }
    \subfigure[ ]
     {  \centering
        \label{fig: sym_real_DAS-CUSUM}%
        \includegraphics[width = .45\textwidth]{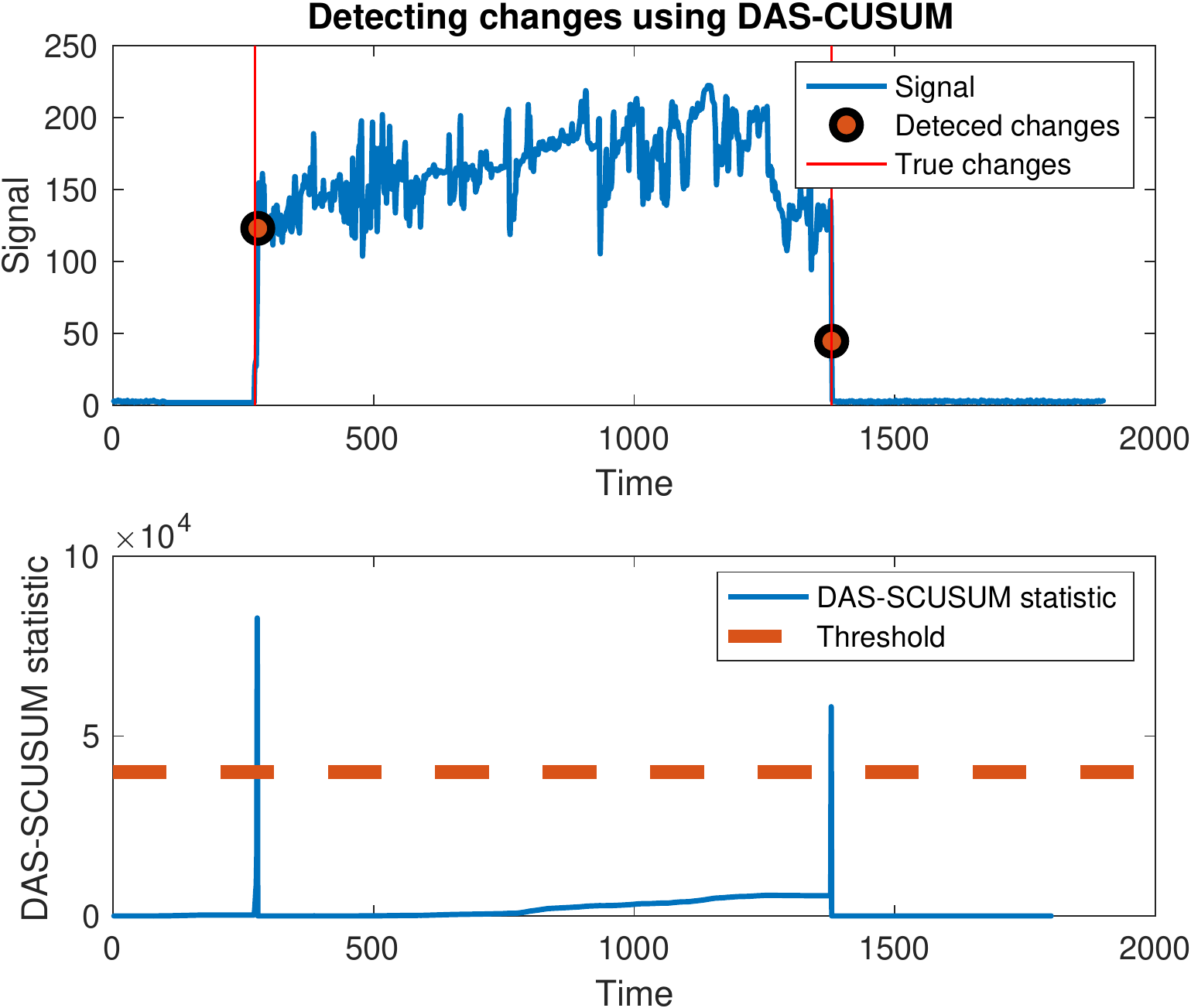}
        \label{fig: symmetry wc}
    }   \caption{Figures \ref{fig: sym_real_CUSUM} shows performance  and \ref{fig: sym_real_DAS-CUSUM} show the advantages of using  DAS-CUSUM for multiple changes over adaptive CUSUM}
   \label{fig: advantage DAS-CUSUM real}
\end{figure}

Due to its symmetric statistic for detecting changes between two distributions, DAS-CUSUM is more useful for detecting multiple changes as compared to GLR and Adaptive CUSUM. This is favorable for detecting multiple changes in real-world problems, as seen in Figure \ref{fig: advantage DAS-CUSUM real}.
which shows readings from a pressure mat that can be seen in Figure \ref{fig: WiSAT}. The mat is inserted beneath a wheelchair cushion and is used to characterize in-seat movement for wheelchair users. When the wheelchair is occupied, the sensor signal has a high mean and variance, whereas when the chair is unoccupied, the signal has a low mean and variance. Detecting changes in occupancy can be treated as a change detection problem. As discussed previously, the asymmetric log-likelihood ratio makes it difficult for both GLR and adaptive CUSUM to detect these changes.

For both Figure \ref{fig: asymmetry wc} and Figure \ref{Fig: Asymmetry real data}, the statistic for getting into the chair (low variance to high variance) is not equal to the statistic for getting out of the chair. For this reason, it is difficult to select a threshold that detects both changes. It can be seen that there is a larger delay in detecting the change while still detecting a false positive change point. Because of the asymmetric statistics, the change for the first statistic is extremely large as compared to the second change. To detect both changes, a lower threshold is set, which causes the first change to be detected really quickly (where the signal is in the middle of the transition). This causes incorrect signal estimates to be used as pre-change estimates causing false change points to be detected. 
Figure \ref{fig: sym_real_DAS-CUSUM} shows the performance of DAS-CUSUM on this signal. The symmetric change statistic provides similar power for detecting both changes without detecting any false positive changes. The symmetric statistic makes it easy to select a threshold to detect multiple changes. This is attractive for real-world scenarios where numerous changes need to be detected when signal changes to unforeseen distributions. 

\begin{figure}[H]%
    \centering
    \subfigure[Adaptive CUSUM misses true change points]
     {  \centering
        \label{fig: wc_CUSUM missed}%
        \includegraphics[width = .45\textwidth]{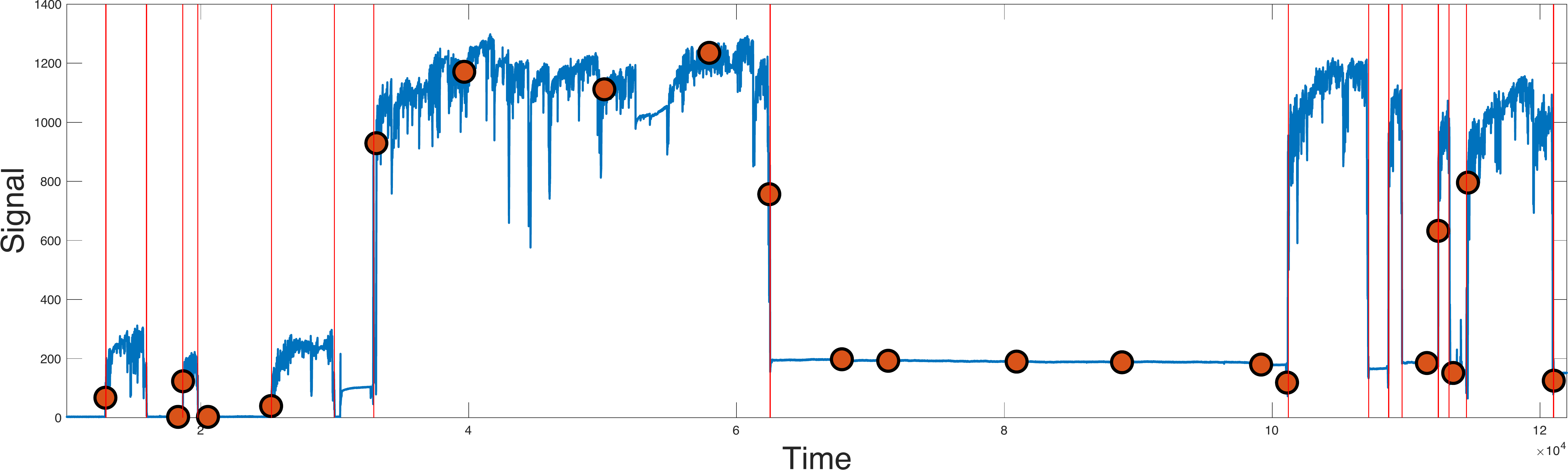}
    }
    \subfigure[GLR  misses true change points]
     {  \centering
        \label{fig: wc_GLR missed}%
        \includegraphics[width = .45\textwidth]{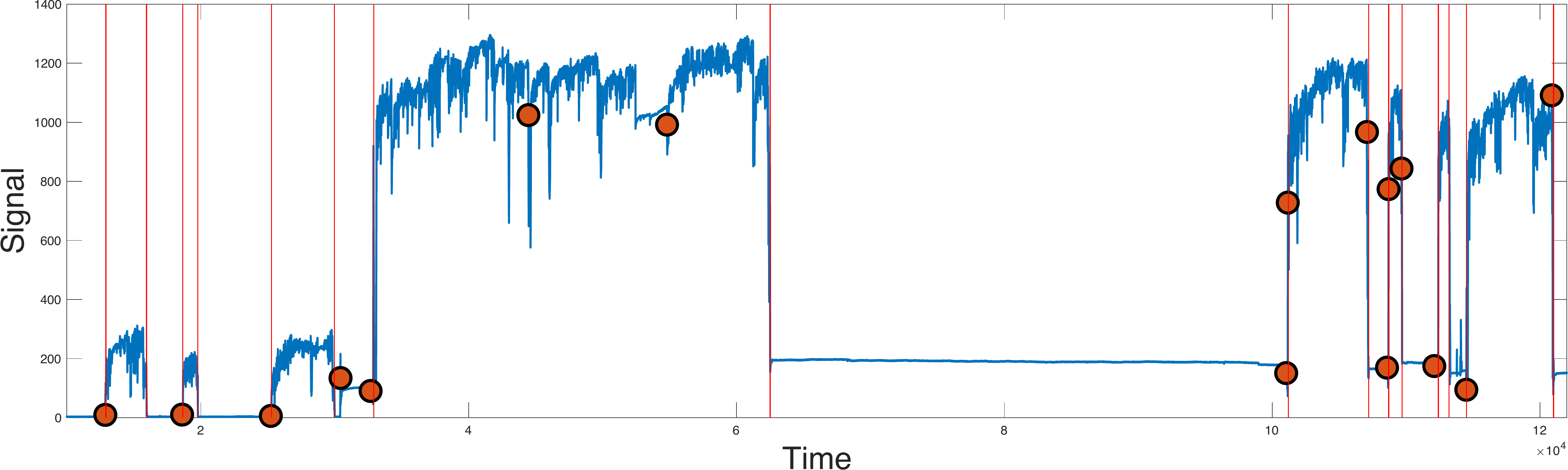}
    }
        \subfigure[Adaptive CUSUM detects false change points]
     {  \centering
        \label{fig: wc_CUSUM false}%
        \includegraphics[width = .45\textwidth]{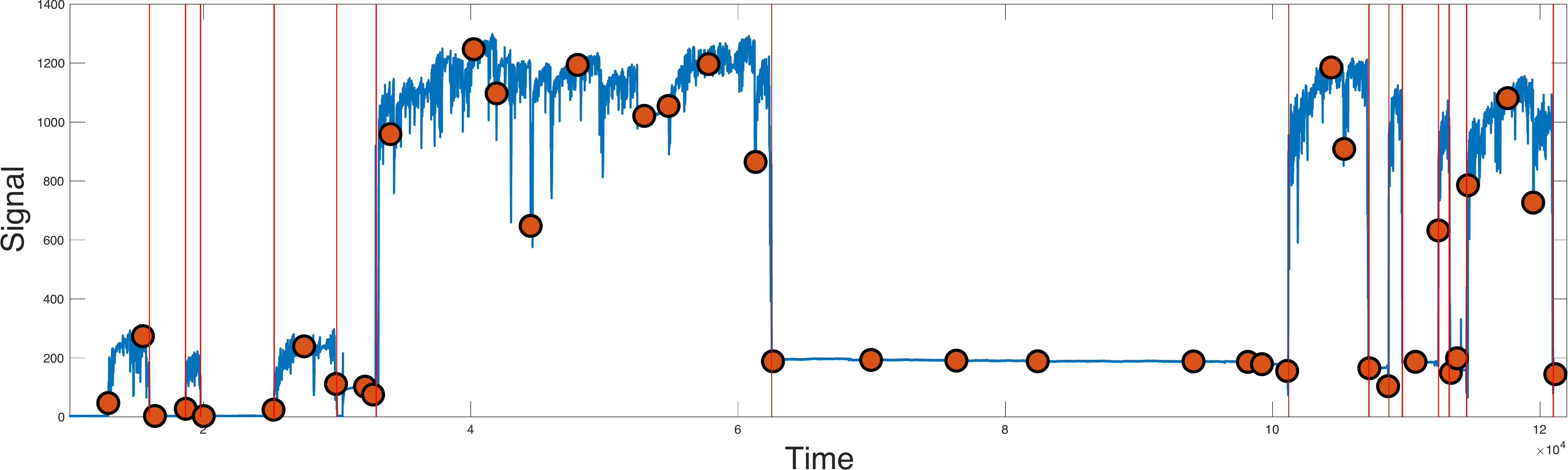}
    }
    \subfigure[GLR detects false change points]
     {  \centering
        \label{fig: wc_GLR false}%
        \includegraphics[width = .45\textwidth]{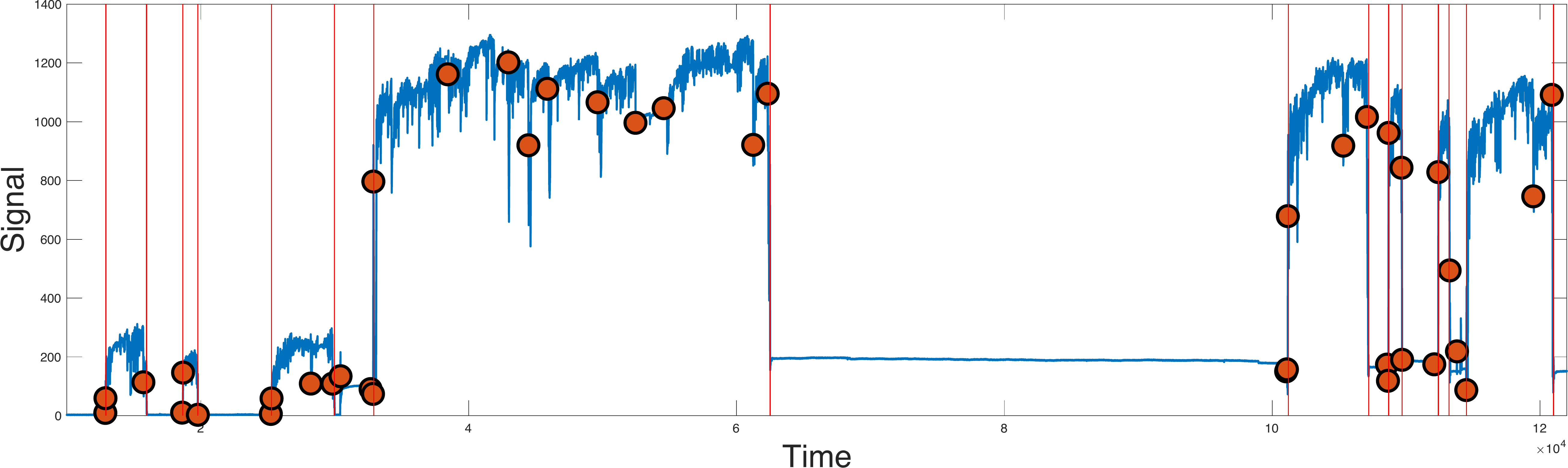}
    }
    \subfigure[DAS-CUSUM correctly detects  true change points ]
     {  \centering
        \label{fig: wc_DAS-CUSUM_correct}%
        \includegraphics[width = .55\textwidth]{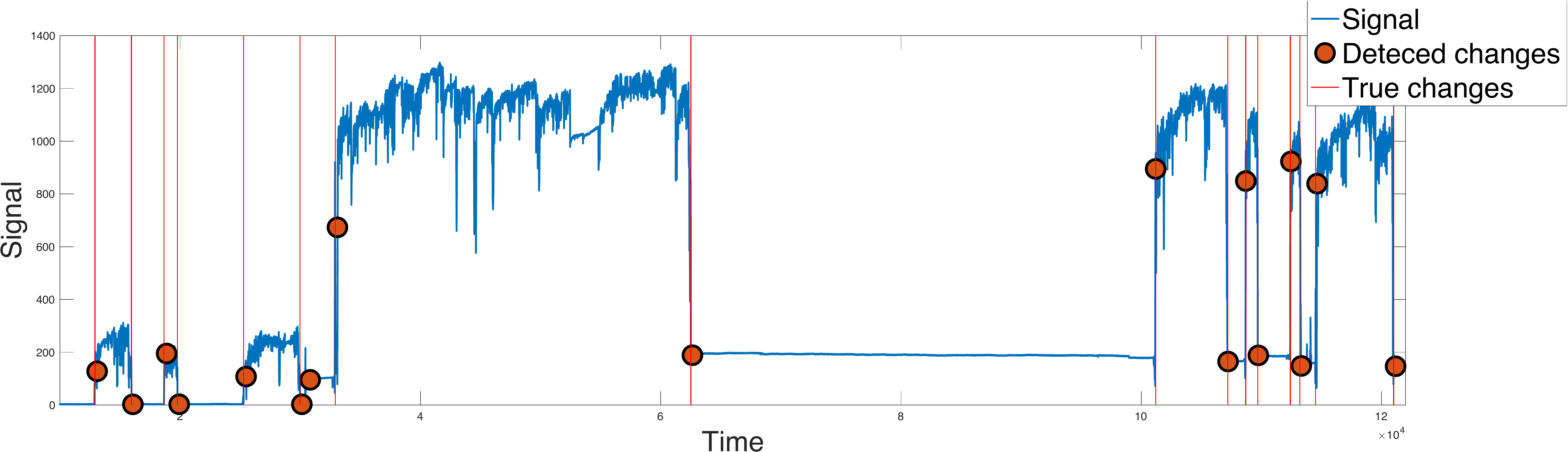}{}
    }   \caption{Comparison of GLR, Adaptive CUSUM, and DAS-CUSUM for detecting multiple change points. The asymmetric log-likelihood ratio makes it difficult for CUSUM and GLR to detect all changes correctly without any false alarms. In Figures \ref{fig: wc_CUSUM missed} and \ref{fig: wc_GLR missed}, a large detection threshold to avoid false change points results in many change missed change points while still detecting a few false change points. Figures \ref{fig: wc_CUSUM false} \ref{fig: wc_GLR false} show how a lower threshold results in many false change points. The symmetric DAS-CUSUM is able to correctly detect all true change points without detecting any false change points} 
   \label{fig: extended_wc}
\end{figure}

Figure \ref{fig: extended_wc} provides an extended example of the occupancy problem. The signal sensors develop drift, and the post-change distribution can change to different unknown distributions at different times. This makes it difficult to use 2-sided CUSUM or other variants as the post-change distribution is not known. In such an example, it can be seen that with symmetric statistics, DAS-CUSUM performs much better than GLR and adaptive CUSUM.  The in-chair distribution is not static. The mean and the variance of the signal changes within the chair, however, these changes are much smaller than the changes in distribution when there is a change in wheelchair occupancy. Symmetric DAS-CUSUM's change statistic is much larger for these occupancy change events, which makes it easy to detect these events without detecting any false alarms. For all methods, a window size of 300 was to estimate the post-change distribution.

\section{Conclusion}

In this work, we have presented DAS-CUSUM, which is a symmetric change point detection procedure. Due to DAS-CUSUM's symmetric incremental statistic, the EDD versus ARL relationship is the same for changes from a distribution $\theta_0$ to  $\theta_1$ and from $\theta_1$ to  $\theta_0$. This symmetric change statistic is helpful when identifying multiple changes in both the mean and variance of a signal. A single threshold can be easily set to detect multiple change points. This is extremely helpful for identifying change points in real-world settings where log-likelihood ratio-based approaches such as  GLR and adaptive CUSUM struggle.  We have derived results that characterize DAS-CUSUM's expected detection delay (EDD)  and average run length (ARL). Extensive simulations are used to validate these results.


%

\bibliography{sequential_analysis.bib}{}
\bibliographystyle{plain}

\section*{Appendix}

\subsection*{A: Proof of Lemma \ref{lemma: v equiv}}
\label{Appendix: A}
\subsubsection*{Computing the inner expectation}

For the expression in~\eqref{eq: Outer expec}, the inner expectation will be first simplified by completing the square to form another normal distribution and then integrating
\begin{align*}
    &\mathbb{E}_{x_t \sim \theta_0}\left[r\left(x_t \right) \Big| \mut, \sigt  \right] \\
    &=\mathbb{E}_{x_t \sim \theta_0}\left[  \exp \left( \deln \left( - \frac{(x_t - \hat{\mu})^2}{2 \hat{\sigma}^2}+\frac{(x_t-\mu_0)^2}{2\sigma_0^2} \right) \right)  \Big| \mut,\sigt\right]\\ &= \int_{-\infty}^{\infty} \frac{1}{\sqrt{2\pi\sigma_0^2}}\exp \left(  - \frac{\deln(x_t - \hat{\mu})^2}{2 \hat{\sigma}^2}+\frac{\deln(x_t-\mu_0)^2}{2\sigma_0^2}  \right)\exp\frac{-(x_t-\mu_0)^2}{2\sigma_0^2}d(x_t)\\
    &= \int_{-\infty}^{\infty} \frac{1}{\sqrt{2\pi\sigma_0^2}} \exp \left(  - \frac{\deln(x_t - \hat{\mu})^2}{2 \hat{\sigma}^2}-\frac{(1 - \deln)(x_t-\mu_0)^2}{2\sigma_0^2}  \right) d(x_t)\\
       &= \int_{-\infty}^{\infty} \frac{1}{\sqrt{2\pi\sigma_0^2}} \exp\left(  - \frac{\deln\sigma_0^2(x_t - \hat{\mu})^2 -(1 - \deln)\sigt^2(x_t-\mu_0)^2}{2\sigt^2\sigma_0^2}  \right) d(x_t)\\
       &= \frac{1}{\sqrt{2\pi\sigma_0^2}} \int_{-\infty}^{\infty}\exp\left( \frac{-(z)\left(x_t^2 - 2x_t\frac{2y}{z} + \frac{y}{z} \right)}{2\sigt^2\sigma_0^2}\right)d(x_t) \\
       &\left(\text{ where: } z  =  \deln\sigma_0^2 + (1 - \deln)\sigt^2 \text{ and } y = \mut\sigma_0^2\deln + \mu_0\sigt^2(1-\deln) \right)\\
       &=\frac{1}{\sqrt{2\pi\sigma_0^2}} \int_{-\infty}^{\infty}\exp\left( \frac{-\left(x^2 - 2x_t\frac{y}{z} + \right(\frac{y}{z}\left)^2 -  \right(\frac{y}{z}\left)^2 +  \frac{y} {z} \right) }{ 2\frac{\sigt^2\sigma_0^2}{z}}\right) d(x_t)\\
       &= \frac{1}{\sqrt{2\pi\sigma_0^2}} \exp\left( -  \frac{\left(\frac{y}{z}\right)^2 + \frac{y}{z}}{2\frac{\sigt^2\sigma_0^2}{z}} \right)\int_{-\infty}^{\infty}\exp\left( \frac{-\left(x_t -\left(\frac{y}{z}\right) \right)^2}{2\underbrace{\frac{\sigt^2\sigma_0^2}{z}}_{\sigma_{12}^2}}\right) d(x_t)\\
       &=\frac{1}{\sqrt{2\pi\sigma_0^2}}\exp\left( \frac{-y^2 + yz}{2z\sigt^2\sigma_0^2} \right) \sqrt{2\pi\sigma_{12}^2}\\
       &=\frac{\sigma_{12}}{\sigma_0}\exp\left(\frac{-\left( \sigt^2\sigma_0^2\deln(1 - \deln)\right) \left( \mu_0^2 - 2\mu_0\mut + \mu_t^2\right) }{2\sigt^2\sigma_0^2(\deln\sigma_0^2 + (1-\deln)\sigt^2)}\right)  \text{  (After substituting $y$ and $z$)}\\
       &= \frac{\sigma_{12}}{\sigma_0}\exp\left( \frac{-(\mu_0 - \mu_t)^2}{\frac{2\sigt^2\sigma_0^2(\deln\sigma_0^2 + (1-\deln)\sigt^2)}{\sigt^2\sigma_0^2\deln(1 - \deln)}} \right) \\
       &= \frac{\sigma_{12}}{\sigma_0}\exp\left(\frac{-(\mu_0 - \mut)^2}{2\frac{\sigma_0^2}{1 - \deln} + 2\frac{\sigt^2}{\deln} } \right),
\end{align*}

where 

    \begin{equation}
       \sigma_{12}^2 = \frac{\sigt^2\sigma_0^2}{\deln\sigma_0^2 + (1 - \deln)\sigt^2}. 
 \label{eq: sig12}
    \end{equation}{}
(Note: $\deln$ should be such that $\sigma_{12}^2  >0$ in~\eqref{eq: sig12}).

Thus 

\begin{align}
     \mathbb{E}_{x_t \sim \theta_0}\left[  \exp \left( \deln \left( - \frac{(x_t - \mut)^2}{2 \sigt^2}+\frac{(x_t-\mu_0)^2}{2\sigma_0^2} \right) \right)  \Big| \mut,\sigt\right] &= \frac{\sigma_{12}}{\sigma_0}\exp\left(\frac{-(\mu_0 - \mut)^2}{2\frac{\sigt^2}{\deln} + 2\frac{\sigma_0^2}{1-\deln}}\right).
     \label{eq: inner expec}
\end{align}

\subsubsection*{Computing the outer expectation}

Plugging in the results of the inner expectation from~\eqref{eq: inner expec} in~\eqref{eq: Outer expec}:
\begin{equation*}
     \mathbb{E}_{\theta_0}\left[\exp({\deln \Tilde{s}_t})\right] = \exp(\deln(-\frac{1}{2} -v))\mathbb{E}_{\mathop{x}_{t+1..w}\sim \theta_0}\left[\exp\left(\deln\left(\frac{\sigma_0^2 + (\mu_0 - \mut)^2}{2\sigt^2} \right) \right) \frac{\sigma_{12}}{\sigma_0}\exp\left(\frac{-(\mu_0 - \mut)^2}{2\frac{\sigt^2}{\deln} + 2\frac{\sigma_0^2}{1-\deln}}\right)\right].
\end{equation*}
Expressing and simplifying the above equation  yields:
\begin{align}
     &\mathbb{E}_{\theta_0}\left[\exp({\deln \Tilde{s}_t})\right] = \exp(\deln(-\frac{1}{2} -v))\mathbb{E}_{\mathop{x}_{t+1..w}\sim \theta_0}\left[\exp\left(\deln\left(\frac{\sigma_0^2 + (\mu_0 - \mut)^2}{2\sigt^2} \right) \right)\frac{\sigma_{12}}{\sigma_0} \exp\left(\frac{-(\mu_0 - \mut)^2}{2\frac{\sigt^2}{\deln} + 2\frac{\sigma_0^2}{1-\deln}}\right)\right] \nonumber \\
     &= \exp(\deln(-\frac{1}{2} -v))\mathbb{E}_{\mathop{x}_{t+1..w}\sim \theta_0}\left[\exp\left( \frac{1}{2}\log\frac{\sigma_{12}^2}{\sigma_0^2} + \deln\left(\frac{\sigma_0^2 + (\mu_0 - \mut)^2}{2\sigt^2} \right)  - \frac{( 1 - \deln)(\deln)(\mu_0 - \mut)^2}{2(1 - \deln)\sigt^2 + 2\deln\sigma_0^2}\right) \right] \nonumber \\
     &= \exp(\deln(-\frac{1}{2} -v))\mathbb{E}_{\mathop{x}_{t+1..w}\sim \theta_0}\left[\exp\left(g(\mut,\sigt^2)\right) \right].
     \label{eq: argument to be subbed for asymptotic}
\end{align}

\subsubsection*{Asymptotic distribution of  $g(\mut,\sigt)$}

Now the asymptotic distribution for the argument of the exponent ($ g(\mut,\sigt)$) within the expectation would be found (when sample mean and sample variance are estimated under the pre-change distribution). This argument is defined as:
\begin{equation}
    g(\mut,\sigt) = \underbrace{\frac{1}{2}\log\left(\frac{\sigt^2}{\deln\sigma_0^2 + (1-\deln)\sigt^2}\right) + \frac{\deln\sigma_0^2}{2\sigt^2}}_{a(\sigt^2)} + \underbrace{\frac{(\mu_0 - \mut)^2}{2}\left(\frac{\deln}{\sigt^2} - \frac{(1-\deln)\deln}{(1-\deln)\sigt^2 + \deln\sigma_0^2}\right)}_{b(\mut,\sigt)}
    \label{eq: g_divided}.
\end{equation}
We now find the distribution of $g(\mut,\sigt)$ when samples samples $x_t..x_{t+w}$ used to calculate $\mut$ and $\sigt$ are distributed by $\theta_0$.
Decomposing $g(\mut,\sigt)$ into two terms:

\begin{equation*}
    a(\sigt^2) = \frac{1}{2}\log\left(\frac{\sigt^2}{\deln\sigma_0^2 + (1-\deln)\sigt^2}\right) + \frac{\deln\sigma_0^2}{2\sigt^2},
\end{equation*}
\begin{equation*}
     b(\mut,\sigt^2) = \frac{(\mu_0 - \mut)^2}{2}\left(\frac{\deln}{\sigt^2} - \frac{(1-\deln)\deln}{(1-\deln)\sigt^2 + \deln\sigma_0^2}\right) .
 \end{equation*}
 
 \subsubsection*{Asymptotic distribution of first term $a(\sigt^2)$}
 
To find the asymptotic distribution of $a(\sigt^2)$, we first recall some results. The asymptotic distribution of sample variance is $\sigt^2$:
\begin{align*}
    \sigt^2 &= \frac{1}{N}\sum_{1=1}^N(x_i - \mut)^2,\\
    \frac{\sigt^2}{\sigma_0^2}        &= \frac{1}{N}\sum_{i=1}^N \underbrace{\frac{(x_i - \mut)^2}{\sigma_0^2}}_{\text{$\chi_1^2$ variables}}. 
\end{align*}
(Note sample variance is divided by $(N-1)$ instead of $N$. Though as $N \xrightarrow{} \infty$, the the sample variance is similar when divided by $N$ or $N-1$. I divide by $N$ to use the tools of central limit theorem which can be found below.)
By the central limit theorem, as $\frac{\sigt^2}{\sigma_0^2} $ is a mean of sum of $\chi_1^2$ variables. These variables have a mean 1 and variance 2:
\begin{align*}
    \sqrt{N}(\frac{\sigt^2}{\sigma_0^2} - 1) \xrightarrow{d} z \sim \mathcal{N}(0,2) .
\end{align*}
Or equivalently
\begin{equation}
     \sqrt{N}(\sigt^2 - \sigma_0^2) \xrightarrow{d} z \sim \mathcal{N}(0,2\sigma_0^4)  .
     \label{eq: samp_var:asymptotic}
\end{equation}
An asymptotically normal estimator $\hat{\theta}$, for the parameter $\theta$, is distributed through:
\begin{equation*}
    \sqrt{n}(\hat{\theta} - \theta) \xrightarrow{d} W \sim \mathcal{N}(0,\sigma^2)
\end{equation*}
For a function $g(\hat{\theta})$, of an asymptotically normal estimator $\hat{\theta}$ of $\theta$,  the delta method states that:
\begin{equation*}
    \sqrt{n}(g(\hat{\theta}) - g(\theta)) \xrightarrow{d} W^* \sim \mathcal{N}(0, g'(\theta)^2\sigma^2)
\end{equation*}
This result is however true only when  $g'(\theta)$ exists and is not 0. Since the sample variance, $\sigt^2$, is asymptotically  normal (as shown in~\eqref{eq: samp_var:asymptotic}), we can try applying the delta method with  $a(\sigt^2)$ in place of $g(\theta)$: 
\begin{equation*}
    \sqrt{n}(a(\sigt^2) - a(\sigma_0^2)) \xrightarrow{d} W^* \sim \mathcal{N}(0, 2(a'(\sigt^2))^2\sigma_0^4)
\end{equation*}
\begin{equation*}
    a'(\sigt^2) = \frac{\deln\sigma_0^2}{-2\sigt^4} + \frac{\deln\sigma_0^2}{2\sigt^2(\deln\sigma_0^2 + (1-\deln)\sigt^2)}
\end{equation*}
\begin{align*}
    a'(\sigma_0^2) &= \frac{\deln\sigma_0^2}{-2\sigma_0^4} + \frac{\deln}{2\sigma_0^2} \\
    &= 0.
\end{align*}

As $a'(\sigma_0^2) = 0$, the delta method cannot be used.
In such a case, the second order delta method can be used if $a''(\sigma_0^2) \neq 0 $ .

\subsubsection*{Second order delta method}

For an asymptotically normal estimator $\hat{\theta}$ for the parameter $\theta$, i.e.,
\begin{equation*}
    \sqrt{n}(\hat{\theta} - \theta) \xrightarrow{d} W \sim \mathcal{N}(0,\sigma^2),
\end{equation*}
the second order delta method \cite{casella2002statistical} states that if there is a function  $g$ on these estimates $\hat{\theta}$, and both $g(\hat{\theta})$ and $g''(\theta_0)$ exist and are non 0, then

\begin{equation*}
    n(g(\hat{\theta}) - g(\theta_0)) \xrightarrow{d} W \sim \sigma_0^2\frac{g''(\sigma_0^2)}{2}\chi_1^2.
\end{equation*}
Since the sample variance, $\sigt^2$, is asymptotically  normal (as shown in~\eqref{eq: samp_var:asymptotic}), we can try applying the second order delta method with  $a(\sigt^2)$ in place of $g(\theta)$ 
\begin{equation}
    n(a(\sigt^2) - a(\sigma_0^2)) \xrightarrow{d} W \sim \sigma_0^4a''(\sigma_0^2)\chi_1^2.
    \label{eq: second_delta_method}
\end{equation}
Finding the double derivative of $a(\sigt^2)$ with respect to $\sigt^2$
\begin{equation*}
        a''(\sigt^2) = \frac{\deln\sigma_0^2}{\sigt^6} -  \frac{\deln\sigma_0^2\sigt^{-2}(1-\deln)+\deln\sigma_0^2\sigt^{-4}(\deln\sigma_0^2 + (1 - \deln)\sigt^2)}{2(\deln\sigma_0^2 + (1-\deln)\sigt^2)^2}. 
\end{equation*}
Plugging in $\sigt$ = $\sigma_0$
\begin{align*}
    a''(\sigma_0^2) &= \frac{\deln}{\sigma_0^4} - \frac{\deln(1-\deln)+\deln}{2\sigma_0^4} \\
    &= \frac{\deln^2}{2\sigma_0^4}.
\end{align*}
Plugging this result in~\eqref{eq: second_delta_method}, we have
\begin{align*}
       n(a(\sigt^2) - a(\sigma_0^2)) \xrightarrow{d} W \sim \frac{\deln^2}{2}\chi_1^2,\\
       a(\sigt^2)  \xrightarrow{d} W^* \sim \frac{\deln^2}{2n}\chi_1^2 + a(\sigma_0^2).
\end{align*}
As $a(\sigma_0^2) = \frac{\deln}{2}$,
\begin{equation}
       a(\sigt^2)  \xrightarrow{d} W^* \sim \frac{\deln^2}{2n}\chi_1^2 + \frac{\deln}{2}
       \label{eq: first term},
\end{equation}
\begin{equation*}
       a(\sigt^2)  \xrightarrow{d} \frac{\deln^2}{2n}z + \frac{\deln}{2} \text{ ( where  } z \sim \chi_1^2).
\end{equation*}
\subsubsection*{Looking at the 2nd term $b(\mut,\sigt^2)$}

The second term is defined as:
\begin{equation*}
    b(\mut,\sigt^2) = \frac{(\mu_0 - \mut)^2}{2}\left(\frac{\deln}{\sigt^2} - \frac{(1-\deln)\deln}{(1-\deln)\sigt^2 + \deln\sigma_0^2}\right). 
\end{equation*}
Manipulating the second term:
\begin{align}
    b(\mut,\sigt^2) &= \frac{(\mu_0 - \mut)^2}{2}\left(\frac{\deln}{\sigt^2} - \frac{(1-\deln)\deln}{(1-\deln)\sigt^2 + \deln\sigma_0^2}\right)  \nonumber \\
    &= \frac{\frac{\sigma_0^2}{N}\frac{(\mu_0 - \mut)^2}{\frac{\sigma_0^2}{N}}}{2}\left(\frac{\deln}{\sigt^2} - \frac{(1-\deln)\deln}{(1-\deln)\sigt^2 + \deln\sigma_0^2}\right) \nonumber \\
    &= \frac{\frac{\sigma_0^2}{N}\frac{(\mu_0 - \mut)^2}{\frac{\sigma_0^2}{N}}}{2}\left(\frac{\deln}{\frac{\sigma_0^2}{N-1}\frac{N-1}{\sigma_0^2}\sigt^2} - \frac{(1-\deln)\deln}{(1-\deln) \frac{\sigma_0^2}{N-1}\frac{N-1}{\sigma_0^2}\sigt^2 + \deln\sigma_0^2}\right) 
    \label{eq: second term to be sub}.
\end{align}
Recall that the distribution of the sample mean and the sample variance are given by:
\begin{align}
    \mut &\sim \mathcal{N}(\mu_0,\frac{\sigma_0^2}{n}) \nonumber ,\\
    \frac{(n-1)}{\sigma_0^2}\sigt^2 &= z \sim \chi_{n-1}^2 .
    \label{eq: dis sample variance}
\end{align}
Also, 
\begin{equation*}
    \frac{(\mu_0 - \mut)^2}{\frac{\sigma_0^2}{n}} =z \sim \chi_{n-1}^2.
\end{equation*}
Using these results in~\eqref{eq: second term to be sub} yields
\begin{equation*}
    b(\mut,\sigt^2) = \frac{\frac{\sigma_0^2}{n}z}{2}\left(\frac{\deln}{\frac{\sigma_0^2}{n-1}z} - \frac{(1-\deln)\deln}{(1-\deln) \frac{\sigma_0^2}{n-1}z + \deln\sigma_0^2}\right). 
\end{equation*}
As $n \xrightarrow{} \infty$, by law of large numbers $\frac{\chi_{n-1}^2}{n-1} \xrightarrow{} 1$. Thus as $n \xrightarrow{} \infty$, $\frac{z}{n-1} \xrightarrow{} 1$ leading to:

\begin{align*}
     b(\mut,\sigt^2) &\xrightarrow{} \frac{\frac{\sigma_0^2}{n}z}{2}\left(\frac{\deln}{\sigma_0^2} - \frac{(1-\deln)\deln}{(1-\deln)\sigma_0^2 + \deln\sigma_0^2}\right)\\
     &= \frac{\deln^2z}{2n}.
\end{align*}

As $z \sim \chi_1^2$, then when $n \xrightarrow{} \infty$,

\begin{equation}
     b(\mut,\sigt^2) \xrightarrow{d} \frac{\deln^2z}{2n} \sim \frac{\deln^2\chi_1^2}{2n}
     \label{eq: second term}.
\end{equation}

\subsubsection*{Combining the two terms}
Note that
\begin{align*}
    g(\mut,\sigt) &= \underbrace{\frac{1}{2}\log\left(\frac{\sigt^2}{\deln\sigma_0^2 + (1-\deln)\sigt^2}\right) + \frac{\deln\sigma_0^2}{2\sigt^2}}_{a(\sigt^2)} + \underbrace{\frac{(\mu_0 - \mut)^2}{2}\left(\frac{\deln}{\sigt^2} - \frac{(1-\deln)\deln}{(1-\deln)\sigt^2 + \deln\sigma_0^2}\right)}_{b(\mut,\sigt)} \\
    &= a(\sigt^2) + b(\mut,\sigt^2).
\end{align*}
Since 
\begin{align*}
    a(\sigt^2)  &\xrightarrow{d} \frac{\deln^2}{2n}z + \frac{\deln}{2},\\
    b(\mut,\sigt^2) &\xrightarrow{d} \frac{\deln^2z}{2N}.
\end{align*}
we have,
\begin{align}
    g(\mut,\sigt^2)  &\xrightarrow{d} \frac{\deln^2}{2n}z + \frac{\deln}{2} + \frac{\deln^2}{2n}z   \text{ (where }  z \sim \chi_1^2) \nonumber \\
    &= \frac{\deln^2}{2n}y + \frac{\deln}{2}  \text{ (where }  y \sim \chi_2^2) .
\end{align}

A chi square variable of $v$ degrees of freedom can be written as a gamma variable, with shape parameter $v/2$, and scale parameter 2. Also if $x \sim \text{Gamma}(a,b)$,  then 
$ k.x \sim \text{Gamma}(a,k.b) $. Thus the asymptotic distribution of $g(\mut,\sigt^2)$ can be written as:
\begin{equation}
      g(\mut,\sigt^2)  \xrightarrow{d} x + \frac{\deln}{2}   \text{ (where }  x \sim \text{Gamma}(a (\text{shape}) = 1, b (\text{scale}) = \frac{\deln^2}{n} ) ) 
      \label{eq: asymotic dist}.
\end{equation}

\subsubsection*{Finding the equivalence factor $\deln$}

The results from~\eqref{eq: asymotic dist} can be used within~\eqref{eq: argument to be subbed for asymptotic} which can be written as:

\begin{equation*}
    \mathbb{E}_{\theta_0}\left[\exp({\deln \Tilde{s}_t})\right] =  \exp(\deln(-\frac{1}{2} -v))\mathbb{E}_{\mathop{x \sim \text{Gamma}(1/2,2\deln^2/n)}}\left[ \exp\left(x +\frac{\deln}{2}\right)\right].
\end{equation*}
Using this result to solve for $\deln$ in the statement of Lemma \ref{lemma: v equiv}
\begin{align}
    \exp(\deln(-\frac{1}{2} -v))\mathbb{E}_{\mathop{x \sim \text{Gamma}(1/2,2\deln^2/n)}}\left[ \exp\left(x +\frac{\deln}{2}\right)\right] &= 1 \nonumber \\
       \exp\left(-\deln v\right) \mathbb{E}_{\mathop{x \sim \text{Gamma}(1/2,2\deln^2/n)}}\underbrace{\left[\exp\left(x\right)\right]}_{\text{MGFunc }} &= 1 \label{eq: before sub in MGF} .
\end{align}
The moment generating function of the gamma distribution is:
\begin{equation*}
     \mathbb{E}_{\mathop{x \sim \text{Gamma}(a,b)}}\left[\exp\left(t x\right)\right] = (1 - tb)^{-a}.
\end{equation*}
Using the moment generating function results in
\begin{align*}
        \exp\left(-\deln v\right)\left(1 - \frac{\deln^2}{n}\right)^{-1} &= 1\\
        \deln v + \log(1 - \frac{\deln^2}{n}) &= 0\\
        v &= \frac{-\log(1 - \frac{\deln^2}{n})}{\deln}  \qed
        \label{eq :obtained v}     
\end{align*}

\end{document}